\documentclass[%
10pt,
reprint,
superscriptaddress,
amsmath,amssymb,
aps,
prb
]{revtex4-1}

\usepackage{amsmath}
\usepackage{graphicx}% Include figure files
\usepackage{dcolumn}% Align table columns on decimal point
\usepackage{bm}% bold math
\usepackage{physics}
\usepackage{braket}

\begin{document}
\title{Multi-mode ultra-strong coupling in circuit quantum electrodynamics}% Force line breaks with \\
%\thanks{A footnote to the article title}%

\author{Sal J. Bosman}
% \altaffiliation[Also at ]{Physics Department, XYZ University.}%Lines break automatically or can be forced with \\
\affiliation{%
Kavli Institute of NanoScience, Delft University of Technology,\\
PO Box 5046, 2600 GA, Delft, The Netherlands.
% This line break forced with \textbackslash\textbackslash
}%
%  \email{s.j.bosman@tudelft.nl}

\author{Mario F. Gely}
% \altaffiliation[Also at ]{Physics Department, XYZ University.}%Lines break automatically or can be forced with \\
\affiliation{%
Kavli Institute of NanoScience, Delft University of Technology,\\
PO Box 5046, 2600 GA, Delft, The Netherlands.
% This line break forced with \textbackslash\textbackslash
}%
%  \email{s.j.bosman@tudelft.nl}
\author{Vibhor Singh}%
 \affiliation{%
Department of Physics, Indian Institute of Science, Bangalore 560012, India
% This line break forced with \textbackslash\textbackslash
}%

\author{Alessandro Bruno}
\affiliation{ 
Qutech Advanced Research Center, Delft University of Technology,\\
Lorentzweg 1, 2628 CJ Delft, The Netherlands.}

\author{Daniel Bothner}%
 \affiliation{%
Kavli Institute of NanoScience, Delft University of Technology,\\
PO Box 5046, 2600 GA, Delft, The Netherlands.
% This line break forced with \textbackslash\textbackslash
}%

\author{Gary A. Steele}

\affiliation{%
Kavli Institute of NanoScience, Delft University of Technology,\\
PO Box 5046, 2600 GA, Delft, The Netherlands.
% This line break forced with \textbackslash\textbackslash
}%

\maketitle

\textbf{{\small
With the introduction of superconducting circuits into the field of quantum optics~\cite{devoret2013superconducting}, many novel experimental demonstrations of the quantum physics of an artificial atom coupled to a single-mode light field have been realized~\cite{schuster_resolving_2007, kirchmair2013observation}. Engineering such quantum systems offers the opportunity to explore extreme regimes of light-matter interaction that are inaccessible with natural systems. For instance the coupling strength $g$ can be increased until it is comparable with the atomic or mode frequency $\omega_{a,m}$~\cite{yoshihara_superconducting_2016,forn-diaz_ultrastrong_2016-1,casanova2010deep} and the atom can be coupled to multiple modes~\cite{sundaresan2015beyond, mckay2015high} which has always challenged our understanding of light-matter interaction \cite{Bethe1947,houck_controlling_2008, filipp2011multimode,arxiv_gely_divergence-free_2017,malekakhlagh2017cutoff}. 
Here, we experimentally realize the first Transmon qubit~\cite{koch2007charge} in the ultra-strong coupling regime, reaching coupling ratios of $g/\omega_{m}=0.19$ and we measure multi-mode interactions through a hybridization of the qubit up to the fifth mode of the resonator. This is enabled by a qubit with 88\% of its capacitance formed by a vacuum-gap capacitance with the center conductor of a coplanar waveguide resonator. In addition to potential applications in quantum information technologies due to its small size and localization of electric fields in vacuum~\cite{cicak2010low}, this new architecture offers the potential to further explore the novel regime of multi-mode ultra-strong coupling.
}}

Superconducting circuits such as microwave cavities and Josephson junction based artificial atoms
\cite{devoret2013superconducting} have opened up a wealth of new experimental possibilities by enabling light-matter coupling that are orders of magnitude stronger than in analogue experiments with natural atoms \cite{raimond2001manipulating} and by taking advantage of the versatility of engineered circuits. 
Experiments such as photon-number resolution~\cite{schuster_resolving_2007} or Schr\"{o}dinger-cat revivals~\cite{kirchmair2013observation} have beautifully displayed the quantum physics of a single-atom coupled to the electromagnetic field of a single mode.
As the field matures, circuits of larger complexity are explored~\cite{forn-diaz_ultrastrong_2016-1,sundaresan2015beyond, mckay2015high}, opening the prospect of controllably studying systems that are theoretically and numerically difficult to understand.

One example is the interaction between an (artificial) atom and an electromagnetic mode where the coupling rate $g$ becomes a considerable fraction to the atomic or mode eigen-frequency $\omega_{a,m}$. This ultra-strong coupling (USC) regime, described by the quantum Rabi model, shows the breakdown of excitation number as conserved quantity, resulting in a significant theoretical challenge~\cite{casanova2010deep,braak2011integrability}.
In the regime of $g/\omega_{a,m} \simeq 1$, known as deep-strong coupling (DSC), a symmetry breaking of the vacuum is predicted\cite{garziano2014vacuum} (\textit{i.e.} qualitative change of the ground state), similar to the Higgs mechanism or Jahn-Teller instability. From a technological standpoint, the USC regime also has potential applications in quantum computation by decreasing gate times~\cite{romero_ultrafast_2012} as well as the performance of quantum memories \cite{arxiv_stassi_quantum_2017}. To date, such experiments have only been realized with flux qubits~\cite{yoshihara_superconducting_2016,forn-diaz_ultrastrong_2016-1} or in the context of digital quantum simulations~\cite{arxiv_langford_experimentally_2016,arxiv_braumuller_analog_2016}. With very strong coupling rates, the additional modes of an electromagnetic resonator become increasingly relevant, and U/DSC can only be understood in these systems if the multi-mode effects are correctly modeled. Previous extensions of the Rabi model have lead to un-physical predictions of dissipation rates~\cite{houck_controlling_2008} or the Lamb shift~\cite{arxiv_gely_divergence-free_2017} arising from a multi-mode interaction. Recently, new models have been developed in which these unphysical predictions no longer arise~\cite{arxiv_gely_divergence-free_2017,malekakhlagh2017cutoff}. However, experiments have yet to reach a parameter regime where such physics becomes relevant.

%
%%%%%%%%%%%%%%%%%%%%%%%%%%%%%%%%%%%%
%%%%%%%%%%%%%%%%%%%%%%%%%%%%%
\begin{figure*}[t!]
\includegraphics{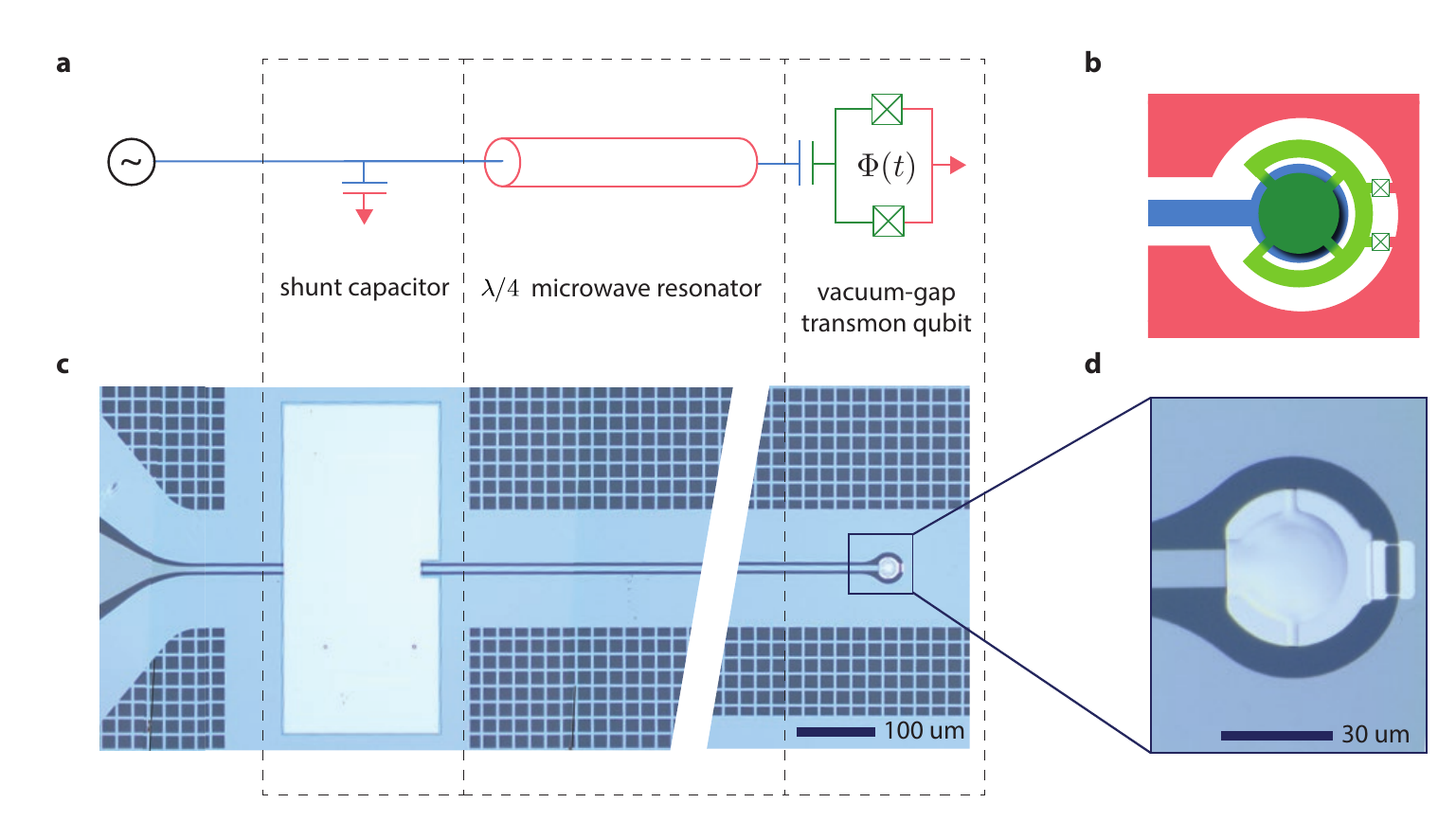}% Here is how to import EPS
                                         % art
\caption{\textbf{Vacuum-gap Transmon circuit architecture. a, }
  Schematic diagram of the equivalent circuit containing a $\lambda/4$ microwave cavity, which on the left is coupled through a shunt capacitor~\cite{bosman2015broadband} to a 50 $\Omega$ port for reflection measurements. On the right, at the voltage anti-node of the resonator, it is capacitively coupled 
  to a Transmon qubit. 
  \textbf{b,} 
  Detailed schematic of the Transmon qubit showing the vacuum-gap capacitor between the center conductor of  the resonator and a suspended superconducting island, which is connected to ground with two Josephson junctions in SQUID geometry (note the matched colors with (a)). 
  \textbf{c,}
  Optical image of a typical device implementing the circuit. 
  \textbf{d,}
  Zoom-in on the qubit showing the suspended capacitor plate above the end of the resonator connected to ground by the junctions.
 }
 \label{fig:device}
\end{figure*}

%%%%%%%%%%%%%%%%%%%%%%%%%%%%%%%%%%%%

Here, we realize a superconducting quantum circuit with a Transmon qubit~\cite{koch2007charge} in the multi-mode USC regime where past extensions standard quantum Rabi models have failed. The qubit consists of a superconducting island shorted to ground by two Josephson junctions in parallel (or SQUID), which is suspended above the voltage anti-node of a quarter wavelength ($\lambda/4$) coplanar waveguide microwave cavity as shown in Figs.~\ref{fig:device}(a,b). This vacuum-gap Transmon architecture offers various possibilities that could prove technologically useful; its an order of magnitude smaller ($30\ \mu$m in diameter) than normal Transmon qubits, its fields are predominantly in vacuum potentially enabling higher coherence~\cite{cicak2010low}, and it offers the possibility to couple \textit{in-situ} to the mechanical motion of the suspended island by applying a voltage bias to the center conductor~\cite{pirkkalainen2013hybrid}. In this study we use this architecture to maximize the coupling. Indeed, the coupling rate is proportional to the capacitance ratio $\beta$ between the qubit capacitance to the resonator $C_c$ and the total qubit capacitance $C_\Sigma$, $\beta=C_c/C_\Sigma$. In this architecture, the vacuum-gap capacitance $C_c$ dominates, leading to $\beta=0.88$. Note that by changing the position of the Transmon along the resonator or its capacitance ratios, its coupling can be reduced to standard coupling rates for other applications.

\begin{figure*}[t!]
\includegraphics[width=16cm]{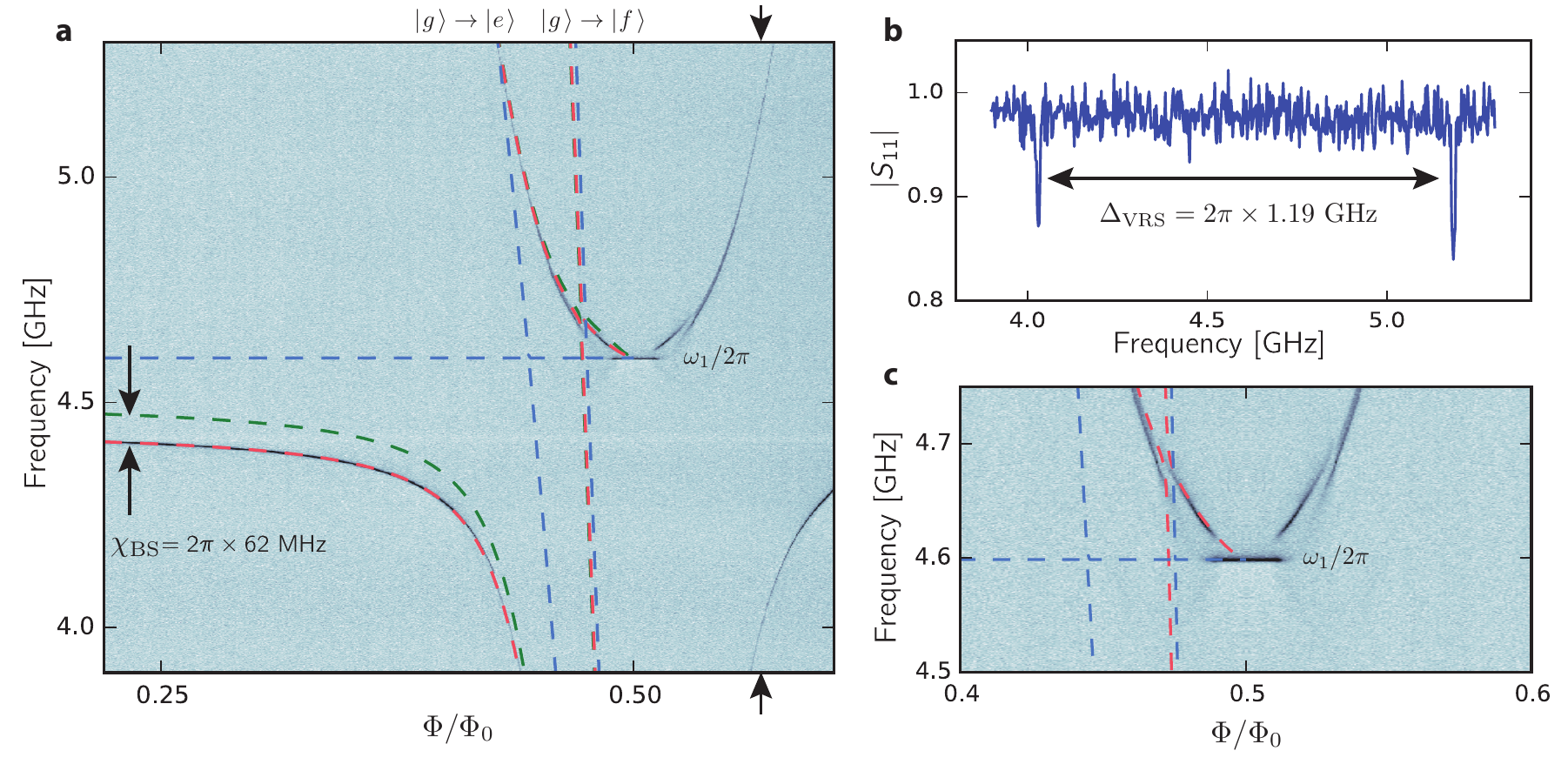}% Here is how to import EPS
\caption{\textbf{Vacuum Rabi splitting. a, }
  The spectral response of device A (155 nm vacuum-gap capacitor) is shown as a function of flux in a single-tone reflection measurement plotted as $|S_{11}|$ (see methods/SI~\cite{SI}). The blue dashed lines indicate the bare (uncoupled) frequency of the fundamental cavity mode, $\omega_1$, and the transition frequencies of the Transmon from the ground state $\ket{g}$, to its first and second excited state, $\ket{e}$ and $\ket{f}$ respectively. The red lines show the hybridized state transitions of the coupled system as fitted from the full spectrum~\cite{SI}. The green lines indicate the dressed state transitions using the rotating wave approximation (RWA) for the same circuit parameters.
%The contribution of the counter-rotating terms to the dispersive shift, the Bloch-Siegert shift, is $\Sigma = 2\pi \times 62$ MHz, about 20 line-width's. 
Note that the splitting is not symmetric with respect to the point at which qubit and mode frequency cross. This is notably due to the renormalization of the charging energy that comes from considering higher resonator modes~\cite{arxiv_gely_divergence-free_2017}.
  \textbf{b,}
  Line-cut showing the vacuum Rabi splitting of the qubit transition with $\omega_1$, resulting in a separation of $\Delta_\text{VRS}=2\pi \times 1.19$ GHz, which is about 281 linewidths of separation.
  \textbf{c,}
  Close up of the spectrum around half a flux quantum (anti-sweet spot), where the qubit frequency is minimal ($\omega_a \lesssim 2\pi \times 1.1$ GHz from the fitted model). There we observe a discrete transition of the spectral response of the circuit towards the bare cavity, $\omega_1$, which we attribute to a decoupling of the qubit and resonator due to a thermally populated qubit. Additionally we observe a small avoided crossing with the $g \leftrightarrow f$ transition.    
  %Line-cut showing the cavity response where the qubit is at the sweetspot. From the fit we extract a total line-width $\kappa=2\pi \times 4.2$ MHz with a coupling coefficient of $\eta=\kappa_{int}/\kappa_{ex}=0.76$, indicating an internal loss rate of $\kappa_{int}=2\pi \times 430$ kHz.
 }
 \label{fig:VRS}
\end{figure*}
\begin{figure*}[t!]
\includegraphics[width=16cm]{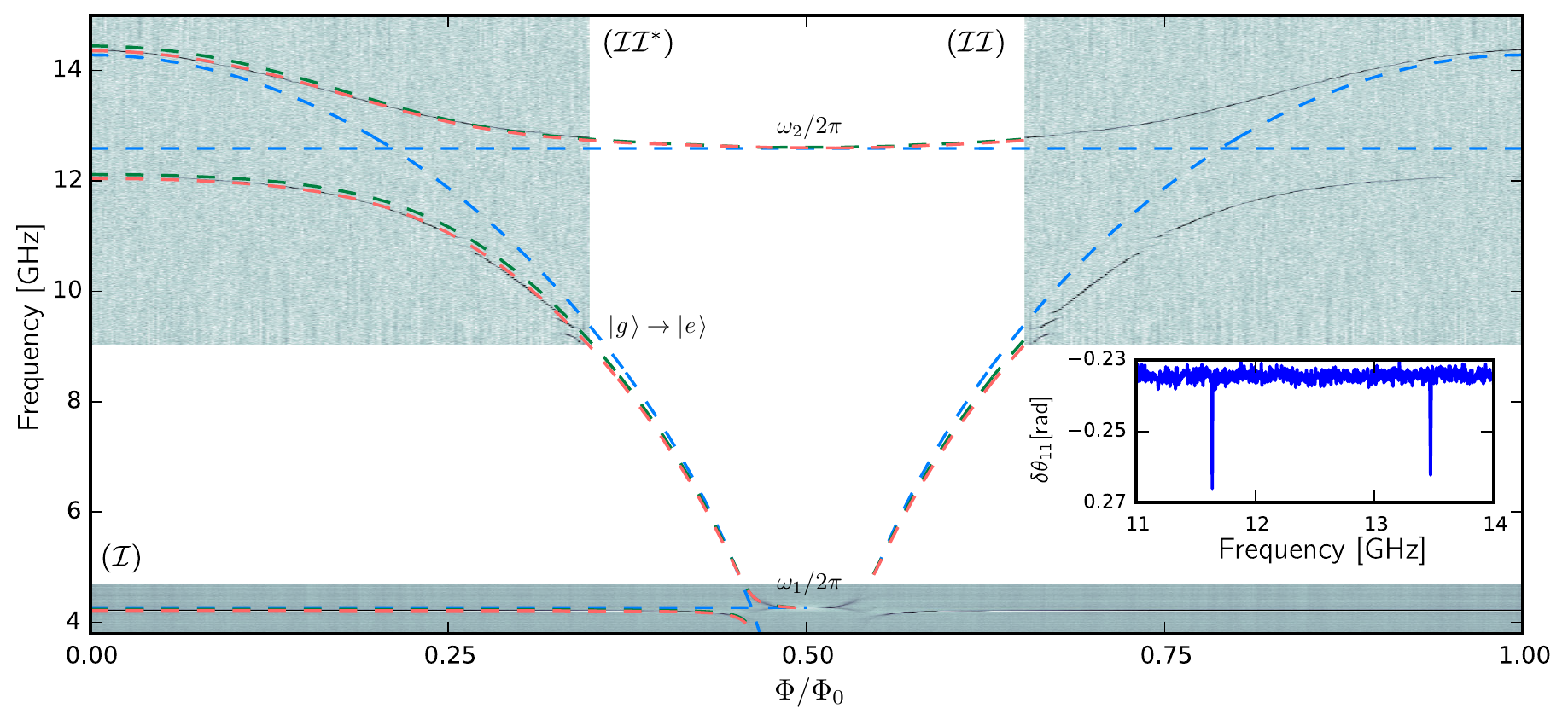}% Here is how to import EPS
\caption{\textbf{Multi-mode spectrum. a, }
 Color plot composed of two spectroscopic measurements of device B (350 nm vacuum-gap capacitor). The bottom frame, $(\mathcal{I})$, shows data obtained as in Fig.~\ref{fig:VRS}. The top frame, $(\mathcal{II})$, shows data from a two-tone spectroscopy measurement from the change in the reflection phase of a weak probe tone at the cavity frequency ($\omega_1$), as a function of a secondary drive tone. We plot the derivative of this phase $\delta\theta_{11}=\partial \varphi[\tilde{S}_{11}(\omega)] /\partial\omega$. Frame $(\mathcal{II}^*)$ is a mirrored copy of the same data.  The blue dashed lines show the transition frequencies of the uncoupled cavities ($\omega_1, \omega_2$) and the first transition ($\ket{g}$ to $\ket{e}$) of the qubit. The red (green) dashed lines show the hybridized spectrum without (with) a RWA. Note that we observe in the top frame a few avoided crossings indicating the presence of two-level systems (TLSs) strongly coupled to the qubit. The inset shows the vacuum Rabi splitting of the qubit with $\omega_2$, giving $\Delta_{2,\text{VRS}}=2\pi \times 1.82$ GHz, plotted as change in reflection angle of the probe tone at $\omega_1$.}

 \label{fig:MM_VRS}    
\end{figure*}

Multi-mode effects play a key role in the physics of this system, but we will start by considering the more simple case of the fundamental mode of the resonator interacting with the Transmon (with levels $\ket{g},\ket{e},\ket{f},...$ of increasing energy). This is described using an extension of the quantum Rabi Hamiltonian~\cite{koch2007charge}
\begin{equation}
  \begin{split} 
  \hat{H}&=\hbar\omega_1 \hat{a}^\dagger\hat{a}
  +\hbar\omega_a(\Phi) \hat{b}^\dagger\hat{b}-\frac{E_c}{2}\hat{b}^\dagger{b}^\dagger\hat{b}\hat{b}\\
  &+\hbar g(\Phi) (\underbrace{\hat{a}\hat{b}^\dagger+\hat{a}^\dagger\hat{b}}_{\mbox{RWA}}+\underbrace{\hat{a}\hat{b}+\hat{a}^\dagger\hat{b}^\dagger}_{\mbox{non-RWA}})\ .
  \end{split}
  \label{eq:hamiltonian}
\end{equation}

Here $\hat{a}$ ($\hat{b}$) is the annihilation operator for the resonator (Transmon) excitations, with frequency $\omega_1$ ($\omega_a(\Phi)$) and $\hbar$ is the reduced Planck constant. The third term introduces the weak anharmonicity of the Transmon, quantified by the charging energy $E_c\simeq e^2/2C_\Sigma$ and the last term describes the coupling of the Transmon to the resonator. Changing the magnetic flux $\Phi$ through the SQUID loop of the Transmon allows us to vary the Josephson energy $E_J(\Phi)$ and hence the frequency $\hbar\omega_a(\Phi)\simeq\sqrt{8E_J(\Phi)E_c}-E_c$ and the coupling $g(\Phi)$\cite{koch2007charge}
\begin{equation}
\hbar g(\Phi)=2e\beta V_\text{zpf} \bigg(\frac{E_J(\Phi)}{32E_c}\bigg)^{1/4},
\end{equation}
with $V_\text{zpf}$ the voltage zero point fluctuations of the microwave cavity and $e$ the electron charge. In our system, USC is due to $\beta=0.88$, whereas $\beta \sim 0.1$ in usual planar geometries. For a Transmon qubit coupled to a single mode, a natural limit on the coupling rate is given by~\cite{bosman_approaching_2017} 
\begin{equation}
  2g<\sqrt{\omega_1\omega_a}\ .
\end{equation}

The light-matter interaction has two types of contributions. 
The first terms conserve excitations, and remain after applying the rotating-wave approximation (RWA). 
The second terms, called counter-rotating terms, add and extract excitations from the qubit and resonator in a pair-wise fashion.  
For sufficiently small couplings the non-RWA terms can be neglected reducing the Rabi model to the Jaynes-Cummings model~\cite{jaynes_comparison_1963}.
For higher couplings the RWA is no longer applicable and the excitation number conservation of the JC model is replaced by a conservation of excitation number parity~\cite{braak2011integrability}. In this regime, making the RWA would lead to a deviation in the energy spectrum of the system known as Bloch-Siegert shift $\chi_\text{BS}$, marking the entry into the USC regime~\cite{forn2010observation}. 

%%%%%%%%%%%%%%%%%%%%%%%%%%%%%%%%%%%

Our samples, depicted in Fig.~\ref{fig:device}(c,d), are fabricated on a sapphire substrate and use as superconductor an alloy of molybdenum-rhenium (MoRe)~\cite{singh2014molybdenum}. 
In a five step electron beam lithography process we pattern the microwave resonator, shunt capacitor dielectric, vacuum-gap sacrificial layer and lift-off mask for the MoRe suspension (see methods for more details). 
In the last step we pattern and deposit the Josephson junctions using aluminum shadow evaporation and perform the release of the vacuum-gap capacitor and aluminum lift-off in the same step.  
In this study we spectroscopically characterize two devices A and B, with vacuum-gap sizes of 155 nm (A) and 350 nm (B).

%%%%%%%%%%%%%%%%%%%%%%%%%%%%%%%%%%%%%%%%%5

Fig.~\ref{fig:VRS} shows the spectral response of device A using single-tone microwave reflectometry at $\sim$14 mK. 
By measuring the complex scattering parameter $S_{11}(\omega)$ of the circuit as a function of an external magnetic field, we can probe the absorption of the circuit at a given frequency within the circulator and amplifier bandwidth of 4-8 GHz (see SI\cite{SI} for full experimental setup).
The transitions of the circuit appear as a dip in the magnitude of the scattering parameter, $|S_{11}|$, thereby mapping the spectrum of the circuit.
From the avoided crossing, depicted in Fig.~\ref{fig:VRS}(a,b), we determine the vacuum Rabi splitting (VRS) to be $\Delta_\text{VRS}=2\pi \times 1.19$ GHz. This provides an estimate of the coupling through the relation $\Delta_\text{VRS}\simeq2g$.
We obtain a ratio $g/\omega_1\simeq \Delta_{\text{VRS}}/2\omega_1=0.13$ indicating we are in the USC regime.

Fig.~\ref{fig:VRS}(c) shows a detailed zoom of the observed spectrum close to half a flux quantum ($\Phi/\Phi_0 \simeq 0.5$).
In this regime, $E_J$ becomes small, such that the qubit frequency goes towards zero for very symmetric junctions, and negligible loop inductance, and the Transmon becomes more like a Cooper-pair box (CPB) as the ratio of $E_J/E_c$ drops~\cite{koch2007charge}. Note that in this regime the physics of the qubit can no longer be described by the Duffing oscillator of Eq.~\ref{eq:hamiltonian}, but rather by the CPB Hamiltonian as was used in all fits of the data~\cite{arxiv_gely_divergence-free_2017}. In this flux region, we observe two notable features.
The first is an anti-crossing at $\Phi/\Phi_0 \sim 0.471$, which we attribute to an avoided crossing with the $\ket{g}$ to $\ket{f}$ transition of the qubit (indicated with the blue dashed line on the right). This shows that in this flux region the qubit behaves like a CPB as such transitions are exponentially suppressed in the Transmon regime~\cite{koch2007charge}. The second feature is a jump of the dressed cavity to the frequency of the bare cavity at $\Phi/\Phi_0 \sim 0.485$.
%The bare cavity frequency can be determined independently in a power dependence measurement.
Such jumps to the bare cavity frequency have been observed before as quantum to classical transitions by applying either high powers of a coherent drive or white noise to the cavity~\cite{bishop2010response,fink2010quantum}. 
In our experiment the critical power of the drive tone that determines the onset of this quasi-harmonic regime is strongly dependent on the flux-bias point. For driving powers corresponding to less than $N_d\sim40$ intra-cavity photons, the bar-like feature becomes power independent, measured with drive powers as low as $N_d \lesssim 0.04$, ruling out a role of the applied drive tone in this feature.
% In the experiment where white noise was applied to artificially elevate the temperature~\cite{fink2010quantum}, the noise power required for the onset of the transition, expressed in terms of photon number, was observed as $N_d \lesssim 1$, which is much lower than typical required coherent drive powers.
We attribute this quenching of the light-matter interaction to a thermal excitation of the low frequency qubit by the environment.
We believe that the transition is observable in our experiment due to a combination of very symmetric junctions in device A, resulting in qubit frequencies that can be excited by the thermal bath of the dilution refrigerator, together with the USC regime that still significantly dresses the cavity resonance even for such large detunings.

%%%%%%%%%%%%%%%%%%%%%%%%%%%%%%%%%%%%%%%%%%%

\begin{figure*}[t!]
\includegraphics[width=16cm]{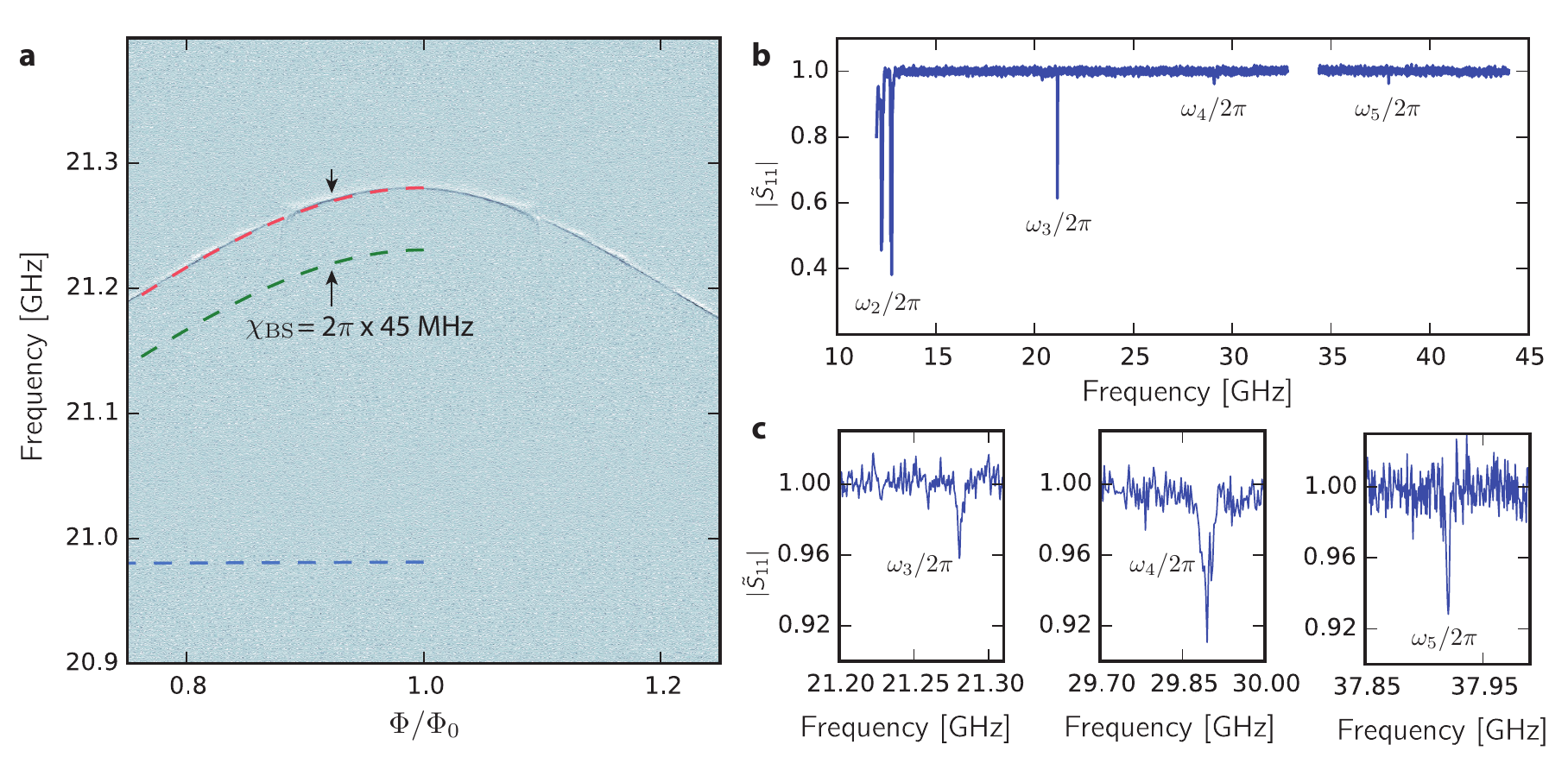}% Here is how to import EPS
\caption{\textbf{Qubit mediated mode-mode interactions. a, }
Two-tone spectroscopy measurement showing the flux dependence of the third mode $\omega_3$ as obtained and shown in Fig.~\ref{fig:MM_VRS}. Due to the strong hybridization over multiple modes, we observe that the qubit-mediated mode-mode coupling is sufficient to observe the effect of driving $\omega_3$ by monitoring the response of the fundamental mode $\omega_1$. The red dashed line indicates the dressed state of $\omega_3$ and the blue dashed line indicates the bare cavity. The green dashed line indicates the predicted line using the RWA, showing the effect of removing the counter-rotating terms, from which we obtain a Bloch-Siegert shift of 45 MHz.
  \textbf{b,} Using the same measurement technique, we show a trace of the normalized reflection coefficient of a weak probe tone positioned at the slope of the resonance of the fundamental mode as a function of a higher frequency drive tone. Here we observe clearly the harmonics of the cavity, including the fourth-mode ($\omega_4$) and fifth-mode ($\omega_5$). The response for the same drive tone power clearly decreases for higher modes as these are further detuned. The data traces measuring $\omega_5$ is $39.5$ dB higher in drive power than the other traces, but making power comparisons is impractical as our microwave measurement setup uses components specified up to 18 GHz. The leftmost peak corresponds to the onset of a frequency region where the system is driven to its linear regime~\cite{bishop2010response} due to the strong drive powers necessary to acquire this data.
  \textbf{c,} Three panels showing a close up of the resonances of $(\omega_3, \omega_4, \omega_5)$, following the harmonics of the fundamental frequency of $\omega_1=2\pi \times 4.268$ GHz.}

 \label{fig:MM}
\end{figure*}

Fig.~\ref{fig:MM_VRS} depicts the spectrum of device B over the full flux periodicity (for device A see SI~\cite{SI}). It is a composition of a single tone measurement, as in Fig.~\ref{fig:VRS}, combined with a two-tone spectroscopy measurement~\cite{wallraff2004strong}. In such a measurement the change in cavity response $\tilde{S}_{11}(\omega)$ is probed using a weak probe tone as a function of a second drive tone at the qubit. Due to the qubit-state dependent dispersive shift of the cavity at $\omega_1$, the reflection of the weak probe tone changes as the drive tone excites the qubit. We observe an avoided crossing of the qubit with the fundamental mode $\omega_1$ ($\Delta_{1,\text{VRS}}=2\pi \times 0.63
$ GHz) and the second mode $\omega_2$, and the frequency maximum of the qubit of $\omega_a = 2\pi \times 14.2$ GHz. From the avoided crossing of the qubit with the second mode we obtain a splitting of $\Delta_{2,\text{VRS}}=2\pi \times 1.82
$ GHz, as shown in the inset. We observe that these two splittings follow the relation $\Delta_{2,\text{VRS}} \simeq 3\Delta_{1,\text{VRS}}$, thereby we observe that the scaling of the VRS evolves linearly with mode number~\cite{filipp2011multimode}. Due to a resolved Bloch-Siegert shift (explained below), we conclude that this device is in the USC regime, wherein the higher modes of the resonator cannot be neglected.

%%%%%%%%%%%%%%%

From the observations of USC to multiple modes in our experiment, it is clear that a quantitative analysis of our experiment should be based on a model that includes multiple modes of the cavity.
Typically this has been done by extending the Rabi model (Eq.~\ref{eq:hamiltonian}) through a square-root increase in coupling strength with mode number $g_m=\sqrt{2m-1}g_0$.
However, as is well established in the literature, such straightforward extensions of the JC and Rabi model to multi-mode systems suffer from divergence problems~\cite{houck_controlling_2008, filipp2011multimode}.
In particular, there is a problem with the predicted qubit frequency due to a divergence of the Lamb shift when the dispersive shift from all of the modes is included~\cite{arxiv_gely_divergence-free_2017}.
In previous experiments where the coupling is small, and the size of the qubit compared to the cavity wave-length is large, a natural cut-off in the number of modes seems to solve these issues and does not reveal the full extent of divergence problems in extended Rabi models~\cite{filipp2011multimode}.
In our case, the small size of our qubit and the USC regime yield a unphysical $~25$ GHz Lamb shift of the qubit following this methodology. Another cut-off associated with the non-zero capacitance of the qubit to ground~\cite{arxiv_gely_divergence-free_2017} leads to a similar shift. This issue can be overcome by using black-box circuit quantization~\cite{nigg2012black}, but with this method we would lose the strict separation of atomic and photonic degrees of freedom typical of the Rabi model, which is essential to estimating the role of counter-rotating terms in the systems spectrum. Additionally, the analysis is then limited to the weakly anharmonic regime of the Transmon qubit whilst our system also enters the Cooper-pair box regime (see Fig.~\ref{fig:VRS}(c)).

%%%%%%%%%%%%%%%

Overcoming this issue led to recent theoretical work~\cite{arxiv_gely_divergence-free_2017}, where a first-principle quantum circuit model was developed based on a lumped element equivalent of this Transmon architecture. This model circumvents the divergence problems of conventional extensions of the Hamiltonian. The red dashed lines in Figs.~\ref{fig:VRS} and \ref{fig:MM_VRS} show a fit of our observed spectrum to the model, demonstrating excellent agreement~\cite{SI}. The fits from the circuit model also allow the extraction of the bare cavity and qubit lines, shown by the blue dashed lines. Note that the definition of the bare qubit frequency strongly differs from typical definitions~\cite{koch2007charge} since it increases (is renormalized) with the number of modes considered in the model. This renormalization is a consequence of the physics of our circuit and compensates the Lamb shift of higher modes. It notably leads to the vacuum Rabi splittings not being symmetrical with respect to the point at which bare qubit and mode cross in Figs.~\ref{fig:VRS} and \ref{fig:MM_VRS}.
An additional feature of the quantum circuit model is that we are able to quantify the relevance of the counter-rotating terms of the interaction between the qubit and the resonator modes.
To do this, we perform the same calculation but removed the counter-rotating terms from the Hamiltonian of the model.
The result is shown by the dashed green lines and allows us to unambiguously extract the resulting vacuum Bloch-Siegert shift $\chi_\text{BS}$, characteristic of the USC regime~\cite{forn2010observation}.
For device A for example, we find a shift of $\chi_\text{BS} = 2\pi \times 62$ MHz (see Fig.~\ref{fig:VRS}), which is about 20 times the cavity line-width, clearly demonstrating our experiment is in the USC regime. Finally we can extract the magnitude of the coupling at its maximum ($\Phi=0$) and obtain for device A a value of $897$ MHz, resulting in a coupling ratio of $g/\omega_1=0.195$

%%%%%%%%%%%%%%%

By examining the composition of the eigenstates obtained from our model, we expect that the qubit should be strongly hybridized with multiple modes of the cavity.
In Fig.~\ref{fig:MM}, we show measurements demonstrating this hybridization.
Using two tone spectroscopy as in Fig.~\ref{fig:MM_VRS}, we are able to observe the higher-modes, by monitoring the response of the hybridized fundamental mode while driving the higher modes.
Fig.~\ref{fig:MM}(a) shows a measurement of the third mode of the cavity $\omega_3$ as a function of flux.
Due to the strong hybridization, we observe a flux tuning of  $\sim$70 MHz despite a detuning from the qubit by $\sim$ 7 GHz.
The red dashed line shows the expected dressed state of $\omega_3$ as predicted from our model, which is in agreement with the data. 
The bare frequency of this mode is 20.98 GHz indicated with the blue dashed line, from which we extract a dispersive shift of 200 - 270 MHz.
Furthermore from the model we find that the counter-rotating terms are crucial for this physics, as the predicted spectrum shifts more than 50 MHz by removing them from the Hamiltonian, as indicated with the green dashed line.   

%%%%%%%%%%%%%%%
Fig.~\ref{fig:MM}(b) shows such a measurement up to 45 GHz. In addition to the third mode shown in Fig.~\ref{fig:MM}(a), we also observe the fourth and the fifth mode of the cavity, demonstrating the qubit induced hybridization over five modes of the cavity extended up to 38 GHz.

To conclude, we have introduced a novel circuit architecture based on the Transmon qubit\cite{koch2007charge}, where a vacuum-gap capacitor significantly dominates the total capacitance of the qubit.
Being ten times smaller than existing Transmon architectures, together with the prospect of higher possible coherence by localizing electric fields in vacuum, this new device could have potential applications in quantum computing technologies. 
Here, we have used this new architecture to maximize the coupling between the qubit and the microwave resonator by increasing the capacitance participation ratio to $\beta\sim0.88$.
Doing so, we realized couplings with the fundamental mode up to $850$ MHz, well within the USC limit, and found that the multi-mode character of the $\lambda/4$ resonator plays a crucial role in the physics of the circuit.
Using a quantum circuit model, we found a Bloch-Siegert shift induced by counter-rotating terms of up to $\chi_\text{BS}=2\pi \times 62$ MHz. 
Combining this architecture with high-impedance microwave resonators~\cite{samkharadze2016high,bosman_approaching_2017} and a smaller free spectral range~\cite{sundaresan2015beyond}, we expect to reach even further into the multi-mode ultra-strong coupling regime to probe exotic states of light and matter~\cite{andersen2016ultrastrong}.\\

\textbf{Methods}

{\small
\textbf{Fabrication}

In the first step, we define the bottom metalization layer of the cavity, including the bottom layers of the shunt-capacitor and the vacuum-gap capacitor, on top of a sapphire substrate.
We use magnetron sputtering to deposit a $45\,$nm thick layer of $60-40$ molybdenum-rhenium (MoRe) alloy and pattern it by means of electron-beam lithography (EBL) and SF$_6$/He reactive ion etching (RIE).
For the definition of the shunt-capacitor dielectric, we deposit a $100\,$nm thick layer of silicon nitride by means of plasma-enhanced chemical vapor deposition and perform the patterning by EBL and wet etching in buffered hydrofluoric acid.
In a third EBL step we pattern the sacrificial layer for the vacuum-gap capacitor, which in our samples consists of a $\sim 160\,$nm thick layer of the electron-beam resist PMGI SF7 diluted $2:1$ with cyclopentanone. 
After the development of the sacrificial layer in L-ethyl-lactate, stopped by rinsing with isopropanol, we reflow the patterned PMGI for $180\,$s at $250$ $^{\circ}$C in order to slightly smooth the stepped edge, facilitating the sidewall metalization in the next step.
The shunt-capacitor and vacuum-gap capacitor top electrodes are fabricated subsequently by means of lift-off technique.
First, we perform EBL to pattern the corresponding PMMA resist layer and secondly, we sputter deposit a $120\,$nm thick layer of MoRe on top.
We do the lift-off in hot xylene, while the sacrificial layer of the vacuum-gap capacitor is not attacked in this process and thus remains unchanged.
In the last step, we fabricate the Josephson junctions using a PMGI/PMMA bilayer lift-off mask, EBL and aluminum shadow-evaporation.
Finally, we perform a simultaneous Al lift-off and the drum release in the resist stripper PRS3000 and dry the sample by means of critical point drying.

\textbf{Data visualization} For the color plots of Figs.~\ref{fig:VRS},\ref{fig:MM_VRS} and \ref{fig:MM}, we applied an image processing filter using Spyview, which histogrammically subtracts the mean of each line of constant frequency with  outlier rejection, 90\% low, 2\% high to remove flux-independent features such as cable resonances. 

\textbf{Author Contributions} S.\@ J.\@ B.\@ and G.\@ A.\@ S.\@ conceived the experiment. S.\@ J.\@ B.\@ designed and fabricated the devices. V.\@ S.\@ , A.\@ B.\@ and G.\@ A.\@ S.\@ provided input for the fabrication. S.\@ J.\@ B.\@ and M.\@ F.\@ G.\@ did the measurements with input of D.\@ B.\@ and G.\@ A.\@ S.\@. M.\@ F.\@ G. and D.\@ B.\@ performed data analysis with input of S.\@ J.\@ B.\@ and G.\@ A.\@ S.\@. Manuscript was written by S.\@ J.\@ B.\@, M.\@ F.\@ G.\@ and G.\@ A.\@ S.\@, and all authors provided comments to the manuscript. G.\@ A.\@ S.\@ supervised the work.

\textbf{Acknowledgments} We wish to acknowledge Enrique Solano, Adrian Parra-Rodriguez and Enrique Rico Ortega for valuable input and discussions.

}

 \bibliography{library}

%merlin.mbs apsrev4-1.bst 2010-07-25 4.21a (PWD, AO, DPC) hacked
%Control: key (0)
%Control: author (8) initials jnrlst
%Control: editor formatted (1) identically to author
%Control: production of article title (-1) disabled
%Control: page (0) single
%Control: year (1) truncated
%Control: production of eprint (0) enabled
\providecommand{\noopsort}[1]{}\providecommand{\singleletter}[1]{#1}%
\begin{thebibliography}{35}%
\makeatletter
\providecommand \@ifxundefined [1]{%
 \@ifx{#1\undefined}
}%
\providecommand \@ifnum [1]{%
 \ifnum #1\expandafter \@firstoftwo
 \else \expandafter \@secondoftwo
 \fi
}%
\providecommand \@ifx [1]{%
 \ifx #1\expandafter \@firstoftwo
 \else \expandafter \@secondoftwo
 \fi
}%
\providecommand \natexlab [1]{#1}%
\providecommand \enquote  [1]{``#1''}%
\providecommand \bibnamefont  [1]{#1}%
\providecommand \bibfnamefont [1]{#1}%
\providecommand \citenamefont [1]{#1}%
\providecommand \href@noop [0]{\@secondoftwo}%
\providecommand \href [0]{\begingroup \@sanitize@url \@href}%
\providecommand \@href[1]{\@@startlink{#1}\@@href}%
\providecommand \@@href[1]{\endgroup#1\@@endlink}%
\providecommand \@sanitize@url [0]{\catcode `\\12\catcode `\$12\catcode
  `\&12\catcode `\#12\catcode `\^12\catcode `\_12\catcode `\%12\relax}%
\providecommand \@@startlink[1]{}%
\providecommand \@@endlink[0]{}%
\providecommand \url  [0]{\begingroup\@sanitize@url \@url }%
\providecommand \@url [1]{\endgroup\@href {#1}{\urlprefix }}%
\providecommand \urlprefix  [0]{URL }%
\providecommand \Eprint [0]{\href }%
\providecommand \doibase [0]{http://dx.doi.org/}%
\providecommand \selectlanguage [0]{\@gobble}%
\providecommand \bibinfo  [0]{\@secondoftwo}%
\providecommand \bibfield  [0]{\@secondoftwo}%
\providecommand \translation [1]{[#1]}%
\providecommand \BibitemOpen [0]{}%
\providecommand \bibitemStop [0]{}%
\providecommand \bibitemNoStop [0]{.\EOS\space}%
\providecommand \EOS [0]{\spacefactor3000\relax}%
\providecommand \BibitemShut  [1]{\csname bibitem#1\endcsname}%
\let\auto@bib@innerbib\@empty
%</preamble>
\bibitem [{\citenamefont {Devoret}\ and\ \citenamefont
  {Schoelkopf}(2013)}]{devoret2013superconducting}%
  \BibitemOpen
  \bibfield  {author} {\bibinfo {author} {\bibfnamefont {M.~H.}\ \bibnamefont
  {Devoret}}\ and\ \bibinfo {author} {\bibfnamefont {R.~J.}\ \bibnamefont
  {Schoelkopf}},\ }\href@noop {} {\bibfield  {journal} {\bibinfo  {journal}
  {Science}\ }\textbf {\bibinfo {volume} {339}},\ \bibinfo {pages} {1169}
  (\bibinfo {year} {2013})}\BibitemShut {NoStop}%
\bibitem [{\citenamefont {Schuster}\ \emph {et~al.}(2007)\citenamefont
  {Schuster}, \citenamefont {Houck}, \citenamefont {Schreier}, \citenamefont
  {Wallraff}, \citenamefont {Gambetta}, \citenamefont {Blais}, \citenamefont
  {Frunzio}, \citenamefont {Majer}, \citenamefont {Johnson}, \citenamefont
  {Devoret}, \citenamefont {Girvin},\ and\ \citenamefont
  {Schoelkopf}}]{schuster_resolving_2007}%
  \BibitemOpen
  \bibfield  {author} {\bibinfo {author} {\bibfnamefont {D.~I.}\ \bibnamefont
  {Schuster}}, \bibinfo {author} {\bibfnamefont {A.~A.}\ \bibnamefont {Houck}},
  \bibinfo {author} {\bibfnamefont {J.~A.}\ \bibnamefont {Schreier}}, \bibinfo
  {author} {\bibfnamefont {A.}~\bibnamefont {Wallraff}}, \bibinfo {author}
  {\bibfnamefont {J.~M.}\ \bibnamefont {Gambetta}}, \bibinfo {author}
  {\bibfnamefont {A.}~\bibnamefont {Blais}}, \bibinfo {author} {\bibfnamefont
  {L.}~\bibnamefont {Frunzio}}, \bibinfo {author} {\bibfnamefont
  {J.}~\bibnamefont {Majer}}, \bibinfo {author} {\bibfnamefont
  {B.}~\bibnamefont {Johnson}}, \bibinfo {author} {\bibfnamefont {M.~H.}\
  \bibnamefont {Devoret}}, \bibinfo {author} {\bibfnamefont {S.~M.}\
  \bibnamefont {Girvin}}, \ and\ \bibinfo {author} {\bibfnamefont {R.~J.}\
  \bibnamefont {Schoelkopf}},\ }\href {\doibase 10.1038/nature05461} {\bibfield
   {journal} {\bibinfo  {journal} {Nature}\ }\textbf {\bibinfo {volume}
  {445}},\ \bibinfo {pages} {515} (\bibinfo {year} {2007})}\BibitemShut
  {NoStop}%
\bibitem [{\citenamefont {Kirchmair}\ \emph {et~al.}(2013)\citenamefont
  {Kirchmair}, \citenamefont {Vlastakis}, \citenamefont {Leghtas},
  \citenamefont {Nigg}, \citenamefont {Paik}, \citenamefont {Ginossar},
  \citenamefont {Mirrahimi}, \citenamefont {Frunzio}, \citenamefont {Girvin},\
  and\ \citenamefont {Schoelkopf}}]{kirchmair2013observation}%
  \BibitemOpen
  \bibfield  {author} {\bibinfo {author} {\bibfnamefont {G.}~\bibnamefont
  {Kirchmair}}, \bibinfo {author} {\bibfnamefont {B.}~\bibnamefont
  {Vlastakis}}, \bibinfo {author} {\bibfnamefont {Z.}~\bibnamefont {Leghtas}},
  \bibinfo {author} {\bibfnamefont {S.~E.}\ \bibnamefont {Nigg}}, \bibinfo
  {author} {\bibfnamefont {H.}~\bibnamefont {Paik}}, \bibinfo {author}
  {\bibfnamefont {E.}~\bibnamefont {Ginossar}}, \bibinfo {author}
  {\bibfnamefont {M.}~\bibnamefont {Mirrahimi}}, \bibinfo {author}
  {\bibfnamefont {L.}~\bibnamefont {Frunzio}}, \bibinfo {author} {\bibfnamefont
  {S.~M.}\ \bibnamefont {Girvin}}, \ and\ \bibinfo {author} {\bibfnamefont
  {R.~J.}\ \bibnamefont {Schoelkopf}},\ }\href@noop {} {\bibfield  {journal}
  {\bibinfo  {journal} {Nature}\ }\textbf {\bibinfo {volume} {495}},\ \bibinfo
  {pages} {205} (\bibinfo {year} {2013})}\BibitemShut {NoStop}%
\bibitem [{\citenamefont {Yoshihara}\ \emph {et~al.}(2016)\citenamefont
  {Yoshihara}, \citenamefont {Fuse}, \citenamefont {Ashhab}, \citenamefont
  {Kakuyanagi}, \citenamefont {Saito},\ and\ \citenamefont
  {Semba}}]{yoshihara_superconducting_2016}%
  \BibitemOpen
  \bibfield  {author} {\bibinfo {author} {\bibfnamefont {F.}~\bibnamefont
  {Yoshihara}}, \bibinfo {author} {\bibfnamefont {T.}~\bibnamefont {Fuse}},
  \bibinfo {author} {\bibfnamefont {S.}~\bibnamefont {Ashhab}}, \bibinfo
  {author} {\bibfnamefont {K.}~\bibnamefont {Kakuyanagi}}, \bibinfo {author}
  {\bibfnamefont {S.}~\bibnamefont {Saito}}, \ and\ \bibinfo {author}
  {\bibfnamefont {K.}~\bibnamefont {Semba}},\ }\href {\doibase
  10.1038/nphys3906} {\bibfield  {journal} {\bibinfo  {journal} {Nature
  Physics}\ }\textbf {\bibinfo {volume} {13}},\ \bibinfo {pages} {44} (\bibinfo
  {year} {2016})}\BibitemShut {NoStop}%
\bibitem [{\citenamefont {Forn-D{\'\i}az}\ \emph {et~al.}(2016)\citenamefont
  {Forn-D{\'\i}az}, \citenamefont {Garc{\'\i}a-Ripoll}, \citenamefont
  {Peropadre}, \citenamefont {Orgiazzi}, \citenamefont {Yurtalan},
  \citenamefont {Belyansky}, \citenamefont {Wilson},\ and\ \citenamefont
  {Lupascu}}]{forn-diaz_ultrastrong_2016-1}%
  \BibitemOpen
  \bibfield  {author} {\bibinfo {author} {\bibfnamefont {P.}~\bibnamefont
  {Forn-D{\'\i}az}}, \bibinfo {author} {\bibfnamefont {J.~J.}\ \bibnamefont
  {Garc{\'\i}a-Ripoll}}, \bibinfo {author} {\bibfnamefont {B.}~\bibnamefont
  {Peropadre}}, \bibinfo {author} {\bibfnamefont {J.-L.}\ \bibnamefont
  {Orgiazzi}}, \bibinfo {author} {\bibfnamefont {M.~A.}\ \bibnamefont
  {Yurtalan}}, \bibinfo {author} {\bibfnamefont {R.}~\bibnamefont {Belyansky}},
  \bibinfo {author} {\bibfnamefont {C.~M.}\ \bibnamefont {Wilson}}, \ and\
  \bibinfo {author} {\bibfnamefont {A.}~\bibnamefont {Lupascu}},\ }\href
  {\doibase 10.1038/nphys3905} {\bibfield  {journal} {\bibinfo  {journal}
  {Nature Phys.}\ }\textbf {\bibinfo {volume} {13}},\ \bibinfo {pages} {39}
  (\bibinfo {year} {2016})}\BibitemShut {NoStop}%
\bibitem [{\citenamefont {Casanova}\ \emph {et~al.}(2010)\citenamefont
  {Casanova}, \citenamefont {Romero}, \citenamefont {Lizuain}, \citenamefont
  {Garc{\'\i}a-Ripoll},\ and\ \citenamefont {Solano}}]{casanova2010deep}%
  \BibitemOpen
  \bibfield  {author} {\bibinfo {author} {\bibfnamefont {J.}~\bibnamefont
  {Casanova}}, \bibinfo {author} {\bibfnamefont {G.}~\bibnamefont {Romero}},
  \bibinfo {author} {\bibfnamefont {I.}~\bibnamefont {Lizuain}}, \bibinfo
  {author} {\bibfnamefont {J.~J.}\ \bibnamefont {Garc{\'\i}a-Ripoll}}, \ and\
  \bibinfo {author} {\bibfnamefont {E.}~\bibnamefont {Solano}},\ }\href@noop {}
  {\bibfield  {journal} {\bibinfo  {journal} {Phys. Rev. Lett.}\ }\textbf
  {\bibinfo {volume} {105}},\ \bibinfo {pages} {263603} (\bibinfo {year}
  {2010})}\BibitemShut {NoStop}%
\bibitem [{\citenamefont {Sundaresan}\ \emph {et~al.}(2015)\citenamefont
  {Sundaresan}, \citenamefont {Liu}, \citenamefont {Sadri}, \citenamefont
  {Sz{\H{o}}cs}, \citenamefont {Underwood}, \citenamefont {Malekakhlagh},
  \citenamefont {T{\"u}reci},\ and\ \citenamefont
  {Houck}}]{sundaresan2015beyond}%
  \BibitemOpen
  \bibfield  {author} {\bibinfo {author} {\bibfnamefont {N.~M.}\ \bibnamefont
  {Sundaresan}}, \bibinfo {author} {\bibfnamefont {Y.}~\bibnamefont {Liu}},
  \bibinfo {author} {\bibfnamefont {D.}~\bibnamefont {Sadri}}, \bibinfo
  {author} {\bibfnamefont {L.~J.}\ \bibnamefont {Sz{\H{o}}cs}}, \bibinfo
  {author} {\bibfnamefont {D.~L.}\ \bibnamefont {Underwood}}, \bibinfo {author}
  {\bibfnamefont {M.}~\bibnamefont {Malekakhlagh}}, \bibinfo {author}
  {\bibfnamefont {H.~E.}\ \bibnamefont {T{\"u}reci}}, \ and\ \bibinfo {author}
  {\bibfnamefont {A.~A.}\ \bibnamefont {Houck}},\ }\href@noop {} {\bibfield
  {journal} {\bibinfo  {journal} {Phys. Rev. X}\ }\textbf {\bibinfo {volume}
  {5}},\ \bibinfo {pages} {021035} (\bibinfo {year} {2015})}\BibitemShut
  {NoStop}%
\bibitem [{\citenamefont {McKay}\ \emph {et~al.}(2015)\citenamefont {McKay},
  \citenamefont {Naik}, \citenamefont {Reinhold}, \citenamefont {Bishop},\ and\
  \citenamefont {Schuster}}]{mckay2015high}%
  \BibitemOpen
  \bibfield  {author} {\bibinfo {author} {\bibfnamefont {D.~C.}\ \bibnamefont
  {McKay}}, \bibinfo {author} {\bibfnamefont {R.}~\bibnamefont {Naik}},
  \bibinfo {author} {\bibfnamefont {P.}~\bibnamefont {Reinhold}}, \bibinfo
  {author} {\bibfnamefont {L.~S.}\ \bibnamefont {Bishop}}, \ and\ \bibinfo
  {author} {\bibfnamefont {D.~I.}\ \bibnamefont {Schuster}},\ }\href@noop {}
  {\bibfield  {journal} {\bibinfo  {journal} {Phys. Rev. Lett.}\ }\textbf
  {\bibinfo {volume} {114}},\ \bibinfo {pages} {080501} (\bibinfo {year}
  {2015})}\BibitemShut {NoStop}%
\bibitem [{\citenamefont {Bethe}(1947)}]{Bethe1947}%
  \BibitemOpen
  \bibfield  {author} {\bibinfo {author} {\bibfnamefont {H.~A.}\ \bibnamefont
  {Bethe}},\ }\href {\doibase 10.1103/PhysRev.73.617} {\bibfield  {journal}
  {\bibinfo  {journal} {Phys. Rev. Lett.}\ }\textbf {\bibinfo {volume} {72}},\
  \bibinfo {pages} {339} (\bibinfo {year} {1947})}\BibitemShut {NoStop}%
\bibitem [{\citenamefont {Houck}\ \emph {et~al.}(2008)\citenamefont {Houck},
  \citenamefont {Schreier}, \citenamefont {Johnson}, \citenamefont {Chow},
  \citenamefont {Koch}, \citenamefont {Gambetta}, \citenamefont {Schuster},
  \citenamefont {Frunzio}, \citenamefont {Devoret}, \citenamefont {Girvin},\
  and\ \citenamefont {Schoelkopf}}]{houck_controlling_2008}%
  \BibitemOpen
  \bibfield  {author} {\bibinfo {author} {\bibfnamefont {A.~A.}\ \bibnamefont
  {Houck}}, \bibinfo {author} {\bibfnamefont {J.~A.}\ \bibnamefont {Schreier}},
  \bibinfo {author} {\bibfnamefont {B.~R.}\ \bibnamefont {Johnson}}, \bibinfo
  {author} {\bibfnamefont {J.~M.}\ \bibnamefont {Chow}}, \bibinfo {author}
  {\bibfnamefont {J.}~\bibnamefont {Koch}}, \bibinfo {author} {\bibfnamefont
  {J.~M.}\ \bibnamefont {Gambetta}}, \bibinfo {author} {\bibfnamefont {D.~I.}\
  \bibnamefont {Schuster}}, \bibinfo {author} {\bibfnamefont {L.}~\bibnamefont
  {Frunzio}}, \bibinfo {author} {\bibfnamefont {M.~H.}\ \bibnamefont
  {Devoret}}, \bibinfo {author} {\bibfnamefont {S.~M.}\ \bibnamefont {Girvin}},
  \ and\ \bibinfo {author} {\bibfnamefont {R.~J.}\ \bibnamefont {Schoelkopf}},\
  }\href@noop {} {\bibfield  {journal} {\bibinfo  {journal} {Phys. Rev. Lett.}\
  }\textbf {\bibinfo {volume} {101}},\ \bibinfo {pages} {080502} (\bibinfo
  {year} {2008})}\BibitemShut {NoStop}%
\bibitem [{\citenamefont {Filipp}\ \emph {et~al.}(2011)\citenamefont {Filipp},
  \citenamefont {G{\"o}ppl}, \citenamefont {Fink}, \citenamefont {Baur},
  \citenamefont {Bianchetti}, \citenamefont {Steffen},\ and\ \citenamefont
  {Wallraff}}]{filipp2011multimode}%
  \BibitemOpen
  \bibfield  {author} {\bibinfo {author} {\bibfnamefont {S.}~\bibnamefont
  {Filipp}}, \bibinfo {author} {\bibfnamefont {M.}~\bibnamefont {G{\"o}ppl}},
  \bibinfo {author} {\bibfnamefont {J.~M.}\ \bibnamefont {Fink}}, \bibinfo
  {author} {\bibfnamefont {M.}~\bibnamefont {Baur}}, \bibinfo {author}
  {\bibfnamefont {R.}~\bibnamefont {Bianchetti}}, \bibinfo {author}
  {\bibfnamefont {L.}~\bibnamefont {Steffen}}, \ and\ \bibinfo {author}
  {\bibfnamefont {A.}~\bibnamefont {Wallraff}},\ }\href@noop {} {\bibfield
  {journal} {\bibinfo  {journal} {Phys. Rev. A}\ }\textbf {\bibinfo {volume}
  {83}},\ \bibinfo {pages} {063827} (\bibinfo {year} {2011})}\BibitemShut
  {NoStop}%
\bibitem [{\citenamefont {Gely}\ \emph {et~al.}(2017)\citenamefont {Gely},
  \citenamefont {Parra-Rodriguez}, \citenamefont {Bothner}, \citenamefont
  {Blanter}, \citenamefont {Bosman}, \citenamefont {Solano},\ and\
  \citenamefont {Steele}}]{arxiv_gely_divergence-free_2017}%
  \BibitemOpen
  \bibfield  {author} {\bibinfo {author} {\bibfnamefont {M.~F.}\ \bibnamefont
  {Gely}}, \bibinfo {author} {\bibfnamefont {A.}~\bibnamefont
  {Parra-Rodriguez}}, \bibinfo {author} {\bibfnamefont {D.}~\bibnamefont
  {Bothner}}, \bibinfo {author} {\bibfnamefont {Y.~M.}\ \bibnamefont
  {Blanter}}, \bibinfo {author} {\bibfnamefont {S.~J.}\ \bibnamefont {Bosman}},
  \bibinfo {author} {\bibfnamefont {E.}~\bibnamefont {Solano}}, \ and\ \bibinfo
  {author} {\bibfnamefont {G.~A.}\ \bibnamefont {Steele}},\ }\href@noop {}
  {\bibfield  {journal} {\bibinfo  {journal} {arXiv:1701.05095}\ } (\bibinfo
  {year} {2017})}\BibitemShut {NoStop}%
\bibitem [{\citenamefont {Malekakhlagh}\ \emph {et~al.}(2017)\citenamefont
  {Malekakhlagh}, \citenamefont {Petrescu},\ and\ \citenamefont
  {T\"{u}reci}}]{malekakhlagh2017cutoff}%
  \BibitemOpen
  \bibfield  {author} {\bibinfo {author} {\bibfnamefont {M.}~\bibnamefont
  {Malekakhlagh}}, \bibinfo {author} {\bibfnamefont {A.}~\bibnamefont
  {Petrescu}}, \ and\ \bibinfo {author} {\bibfnamefont {H.~E.}\ \bibnamefont
  {T\"{u}reci}},\ }\href@noop {} {\bibfield  {journal} {\bibinfo  {journal}
  {arXiv:1701.07935v2}\ } (\bibinfo {year} {2017})}\BibitemShut {NoStop}%
\bibitem [{\citenamefont {Koch}\ \emph {et~al.}(2007)\citenamefont {Koch},
  \citenamefont {Yu}, \citenamefont {Gambetta}, \citenamefont {Houck},
  \citenamefont {Schuster}, \citenamefont {Majer}, \citenamefont {Blais},
  \citenamefont {Devoret}, \citenamefont {Girvin},\ and\ \citenamefont
  {Schoelkopf}}]{koch2007charge}%
  \BibitemOpen
  \bibfield  {author} {\bibinfo {author} {\bibfnamefont {J.}~\bibnamefont
  {Koch}}, \bibinfo {author} {\bibfnamefont {T.~M.}\ \bibnamefont {Yu}},
  \bibinfo {author} {\bibfnamefont {J.}~\bibnamefont {Gambetta}}, \bibinfo
  {author} {\bibfnamefont {A.~A.}\ \bibnamefont {Houck}}, \bibinfo {author}
  {\bibfnamefont {D.~I.}\ \bibnamefont {Schuster}}, \bibinfo {author}
  {\bibfnamefont {J.}~\bibnamefont {Majer}}, \bibinfo {author} {\bibfnamefont
  {A.}~\bibnamefont {Blais}}, \bibinfo {author} {\bibfnamefont {M.~H.}\
  \bibnamefont {Devoret}}, \bibinfo {author} {\bibfnamefont {S.~M.}\
  \bibnamefont {Girvin}}, \ and\ \bibinfo {author} {\bibfnamefont {R.~J.}\
  \bibnamefont {Schoelkopf}},\ }\href@noop {} {\bibfield  {journal} {\bibinfo
  {journal} {Phys. Rev. A}\ }\textbf {\bibinfo {volume} {76}},\ \bibinfo
  {pages} {042319} (\bibinfo {year} {2007})}\BibitemShut {NoStop}%
\bibitem [{\citenamefont {Cicak}\ \emph {et~al.}(2010)\citenamefont {Cicak},
  \citenamefont {Li}, \citenamefont {Strong}, \citenamefont {Allman},
  \citenamefont {Altomare}, \citenamefont {Sirois}, \citenamefont {Whittaker},
  \citenamefont {Teufel},\ and\ \citenamefont {Simmonds}}]{cicak2010low}%
  \BibitemOpen
  \bibfield  {author} {\bibinfo {author} {\bibfnamefont {K.}~\bibnamefont
  {Cicak}}, \bibinfo {author} {\bibfnamefont {D.}~\bibnamefont {Li}}, \bibinfo
  {author} {\bibfnamefont {J.~A.}\ \bibnamefont {Strong}}, \bibinfo {author}
  {\bibfnamefont {M.~S.}\ \bibnamefont {Allman}}, \bibinfo {author}
  {\bibfnamefont {F.}~\bibnamefont {Altomare}}, \bibinfo {author}
  {\bibfnamefont {A.~J.}\ \bibnamefont {Sirois}}, \bibinfo {author}
  {\bibfnamefont {J.~D.}\ \bibnamefont {Whittaker}}, \bibinfo {author}
  {\bibfnamefont {J.~D.}\ \bibnamefont {Teufel}}, \ and\ \bibinfo {author}
  {\bibfnamefont {R.~W.}\ \bibnamefont {Simmonds}},\ }\href@noop {} {\bibfield
  {journal} {\bibinfo  {journal} {Appl. Phys. Lett.}\ }\textbf {\bibinfo
  {volume} {96}},\ \bibinfo {pages} {093502} (\bibinfo {year}
  {2010})}\BibitemShut {NoStop}%
\bibitem [{\citenamefont {Raimond}\ \emph {et~al.}(2001)\citenamefont
  {Raimond}, \citenamefont {Brune},\ and\ \citenamefont
  {Haroche}}]{raimond2001manipulating}%
  \BibitemOpen
  \bibfield  {author} {\bibinfo {author} {\bibfnamefont {J.-M.}\ \bibnamefont
  {Raimond}}, \bibinfo {author} {\bibfnamefont {M.}~\bibnamefont {Brune}}, \
  and\ \bibinfo {author} {\bibfnamefont {S.}~\bibnamefont {Haroche}},\
  }\href@noop {} {\bibfield  {journal} {\bibinfo  {journal} {Rev. Mod. Phys.}\
  }\textbf {\bibinfo {volume} {73}},\ \bibinfo {pages} {565} (\bibinfo {year}
  {2001})}\BibitemShut {NoStop}%
\bibitem [{\citenamefont {Braak}(2011)}]{braak2011integrability}%
  \BibitemOpen
  \bibfield  {author} {\bibinfo {author} {\bibfnamefont {D.}~\bibnamefont
  {Braak}},\ }\href@noop {} {\bibfield  {journal} {\bibinfo  {journal} {Phys.
  Rev. Lett.}\ }\textbf {\bibinfo {volume} {107}},\ \bibinfo {pages} {100401}
  (\bibinfo {year} {2011})}\BibitemShut {NoStop}%
\bibitem [{\citenamefont {Garziano}\ \emph {et~al.}(2014)\citenamefont
  {Garziano}, \citenamefont {Stassi}, \citenamefont {Ridolfo}, \citenamefont
  {Di~Stefano},\ and\ \citenamefont {Savasta}}]{garziano2014vacuum}%
  \BibitemOpen
  \bibfield  {author} {\bibinfo {author} {\bibfnamefont {L.}~\bibnamefont
  {Garziano}}, \bibinfo {author} {\bibfnamefont {R.}~\bibnamefont {Stassi}},
  \bibinfo {author} {\bibfnamefont {A.}~\bibnamefont {Ridolfo}}, \bibinfo
  {author} {\bibfnamefont {O.}~\bibnamefont {Di~Stefano}}, \ and\ \bibinfo
  {author} {\bibfnamefont {S.}~\bibnamefont {Savasta}},\ }\href@noop {}
  {\bibfield  {journal} {\bibinfo  {journal} {Phys. Rev. A}\ }\textbf {\bibinfo
  {volume} {90}},\ \bibinfo {pages} {043817} (\bibinfo {year}
  {2014})}\BibitemShut {NoStop}%
\bibitem [{\citenamefont {Romero}\ \emph {et~al.}(2012)\citenamefont {Romero},
  \citenamefont {Ballester}, \citenamefont {Wang}, \citenamefont {Scarani},\
  and\ \citenamefont {Solano}}]{romero_ultrafast_2012}%
  \BibitemOpen
  \bibfield  {author} {\bibinfo {author} {\bibfnamefont {G.}~\bibnamefont
  {Romero}}, \bibinfo {author} {\bibfnamefont {D.}~\bibnamefont {Ballester}},
  \bibinfo {author} {\bibfnamefont {Y.~M.}\ \bibnamefont {Wang}}, \bibinfo
  {author} {\bibfnamefont {V.}~\bibnamefont {Scarani}}, \ and\ \bibinfo
  {author} {\bibfnamefont {E.}~\bibnamefont {Solano}},\ }\href {\doibase
  10.1103/PhysRevLett.108.120501} {\bibfield  {journal} {\bibinfo  {journal}
  {Phys. Rev. Lett.}\ }\textbf {\bibinfo {volume} {108}},\ \bibinfo {pages}
  {120501} (\bibinfo {year} {2012})}\BibitemShut {NoStop}%
\bibitem [{\citenamefont {Stassi}\ and\ \citenamefont
  {Nori}(2017)}]{arxiv_stassi_quantum_2017}%
  \BibitemOpen
  \bibfield  {author} {\bibinfo {author} {\bibfnamefont {R.}~\bibnamefont
  {Stassi}}\ and\ \bibinfo {author} {\bibfnamefont {F.}~\bibnamefont {Nori}},\
  }\href@noop {} {\bibfield  {journal} {\bibinfo  {journal} {arXiv:1703.08951}\
  } (\bibinfo {year} {2017})}\BibitemShut {NoStop}%
\bibitem [{\citenamefont {Langford}\ \emph {et~al.}(2016)\citenamefont
  {Langford}, \citenamefont {Sagastizabal}, \citenamefont {Kounalakis},
  \citenamefont {Dickel}, \citenamefont {Bruno}, \citenamefont {Luthi},
  \citenamefont {Thoen}, \citenamefont {Endo},\ and\ \citenamefont
  {DiCarlo}}]{arxiv_langford_experimentally_2016}%
  \BibitemOpen
  \bibfield  {author} {\bibinfo {author} {\bibfnamefont {N.~K.}\ \bibnamefont
  {Langford}}, \bibinfo {author} {\bibfnamefont {R.}~\bibnamefont
  {Sagastizabal}}, \bibinfo {author} {\bibfnamefont {M.}~\bibnamefont
  {Kounalakis}}, \bibinfo {author} {\bibfnamefont {C.}~\bibnamefont {Dickel}},
  \bibinfo {author} {\bibfnamefont {A.}~\bibnamefont {Bruno}}, \bibinfo
  {author} {\bibfnamefont {F.}~\bibnamefont {Luthi}}, \bibinfo {author}
  {\bibfnamefont {D.~J.}\ \bibnamefont {Thoen}}, \bibinfo {author}
  {\bibfnamefont {A.}~\bibnamefont {Endo}}, \ and\ \bibinfo {author}
  {\bibfnamefont {L.}~\bibnamefont {DiCarlo}},\ }\href@noop {} {\bibfield
  {journal} {\bibinfo  {journal} {arXiv:1610.10065}\ } (\bibinfo {year}
  {2016})}\BibitemShut {NoStop}%
\bibitem [{\citenamefont {Braum{\"u}ller}\ \emph {et~al.}(2016)\citenamefont
  {Braum{\"u}ller}, \citenamefont {Marthaler}, \citenamefont {Schneider},
  \citenamefont {Stehli}, \citenamefont {Rotzinger}, \citenamefont {Weides},\
  and\ \citenamefont {Ustinov}}]{arxiv_braumuller_analog_2016}%
  \BibitemOpen
  \bibfield  {author} {\bibinfo {author} {\bibfnamefont {J.}~\bibnamefont
  {Braum{\"u}ller}}, \bibinfo {author} {\bibfnamefont {M.}~\bibnamefont
  {Marthaler}}, \bibinfo {author} {\bibfnamefont {A.}~\bibnamefont
  {Schneider}}, \bibinfo {author} {\bibfnamefont {A.}~\bibnamefont {Stehli}},
  \bibinfo {author} {\bibfnamefont {H.}~\bibnamefont {Rotzinger}}, \bibinfo
  {author} {\bibfnamefont {M.}~\bibnamefont {Weides}}, \ and\ \bibinfo {author}
  {\bibfnamefont {A.~V.}\ \bibnamefont {Ustinov}},\ }\href@noop {} {\bibfield
  {journal} {\bibinfo  {journal} {arXiv:1611.08404}\ } (\bibinfo {year}
  {2016})}\BibitemShut {NoStop}%
\bibitem [{\citenamefont {Nigg}\ \emph {et~al.}(2012)\citenamefont {Nigg},
  \citenamefont {Paik}, \citenamefont {Vlastakis}, \citenamefont {Kirchmair},
  \citenamefont {Shankar}, \citenamefont {Frunzio}, \citenamefont {Devoret},
  \citenamefont {Schoelkopf},\ and\ \citenamefont {Girvin}}]{nigg2012black}%
  \BibitemOpen
  \bibfield  {author} {\bibinfo {author} {\bibfnamefont {S.~E.}\ \bibnamefont
  {Nigg}}, \bibinfo {author} {\bibfnamefont {H.}~\bibnamefont {Paik}}, \bibinfo
  {author} {\bibfnamefont {B.}~\bibnamefont {Vlastakis}}, \bibinfo {author}
  {\bibfnamefont {G.}~\bibnamefont {Kirchmair}}, \bibinfo {author}
  {\bibfnamefont {S.}~\bibnamefont {Shankar}}, \bibinfo {author} {\bibfnamefont
  {L.}~\bibnamefont {Frunzio}}, \bibinfo {author} {\bibfnamefont {M.~H.}\
  \bibnamefont {Devoret}}, \bibinfo {author} {\bibfnamefont {R.~J.}\
  \bibnamefont {Schoelkopf}}, \ and\ \bibinfo {author} {\bibfnamefont {S.~M.}\
  \bibnamefont {Girvin}},\ }\href@noop {} {\bibfield  {journal} {\bibinfo
  {journal} {Phys. Rev. Lett.}\ }\textbf {\bibinfo {volume} {108}},\ \bibinfo
  {pages} {240502} (\bibinfo {year} {2012})}\BibitemShut {NoStop}%
\bibitem [{\citenamefont {Bosman}\ \emph {et~al.}(2015)\citenamefont {Bosman},
  \citenamefont {Singh}, \citenamefont {Bruno},\ and\ \citenamefont
  {Steele}}]{bosman2015broadband}%
  \BibitemOpen
  \bibfield  {author} {\bibinfo {author} {\bibfnamefont {S.~J.}\ \bibnamefont
  {Bosman}}, \bibinfo {author} {\bibfnamefont {V.}~\bibnamefont {Singh}},
  \bibinfo {author} {\bibfnamefont {A.}~\bibnamefont {Bruno}}, \ and\ \bibinfo
  {author} {\bibfnamefont {G.~A.}\ \bibnamefont {Steele}},\ }\href@noop {}
  {\bibfield  {journal} {\bibinfo  {journal} {Appl. Phys. Lett.}\ }\textbf
  {\bibinfo {volume} {107}},\ \bibinfo {pages} {192602} (\bibinfo {year}
  {2015})}\BibitemShut {NoStop}%
\bibitem [{\citenamefont {Pirkkalainen}\ \emph {et~al.}(2013)\citenamefont
  {Pirkkalainen}, \citenamefont {Cho}, \citenamefont {Li}, \citenamefont
  {Paraoanu}, \citenamefont {Hakonen},\ and\ \citenamefont
  {Sillanp{\"a}{\"a}}}]{pirkkalainen2013hybrid}%
  \BibitemOpen
  \bibfield  {author} {\bibinfo {author} {\bibfnamefont {J.-M.}\ \bibnamefont
  {Pirkkalainen}}, \bibinfo {author} {\bibfnamefont {S.}~\bibnamefont {Cho}},
  \bibinfo {author} {\bibfnamefont {J.}~\bibnamefont {Li}}, \bibinfo {author}
  {\bibfnamefont {G.}~\bibnamefont {Paraoanu}}, \bibinfo {author}
  {\bibfnamefont {P.}~\bibnamefont {Hakonen}}, \ and\ \bibinfo {author}
  {\bibfnamefont {M.}~\bibnamefont {Sillanp{\"a}{\"a}}},\ }\href@noop {}
  {\bibfield  {journal} {\bibinfo  {journal} {Nature}\ }\textbf {\bibinfo
  {volume} {494}},\ \bibinfo {pages} {211} (\bibinfo {year}
  {2013})}\BibitemShut {NoStop}%
\bibitem [{SI()}]{SI}%
  \BibitemOpen
  \href@noop {} {}\bibinfo {note} {See supplementary material.}\BibitemShut
  {Stop}%
\bibitem [{\citenamefont {Bosman}\ \emph {et~al.}(2017)\citenamefont {Bosman},
  \citenamefont {Gely}, \citenamefont {Singh}, \citenamefont {Bothner},
  \citenamefont {Castellanos-Gomez},\ and\ \citenamefont
  {Steele}}]{bosman_approaching_2017}%
  \BibitemOpen
  \bibfield  {author} {\bibinfo {author} {\bibfnamefont {S.~J.}\ \bibnamefont
  {Bosman}}, \bibinfo {author} {\bibfnamefont {M.~F.}\ \bibnamefont {Gely}},
  \bibinfo {author} {\bibfnamefont {V.}~\bibnamefont {Singh}}, \bibinfo
  {author} {\bibfnamefont {D.}~\bibnamefont {Bothner}}, \bibinfo {author}
  {\bibfnamefont {A.}~\bibnamefont {Castellanos-Gomez}}, \ and\ \bibinfo
  {author} {\bibfnamefont {G.~A.}\ \bibnamefont {Steele}},\ }\href@noop {}
  {\bibfield  {journal} {\bibinfo  {journal} {arXiv:1704.04421}\ } (\bibinfo
  {year} {2017})},\ \Eprint {http://arxiv.org/abs/1704.04421}
  {arXiv:1704.04421} \BibitemShut {NoStop}%
\bibitem [{\citenamefont {Jaynes}\ and\ \citenamefont
  {Cummings}(1963)}]{jaynes_comparison_1963}%
  \BibitemOpen
  \bibfield  {author} {\bibinfo {author} {\bibfnamefont {E.~T.}\ \bibnamefont
  {Jaynes}}\ and\ \bibinfo {author} {\bibfnamefont {F.~W.}\ \bibnamefont
  {Cummings}},\ }\href {\doibase 10.1109/PROC.1963.1664} {\bibfield  {journal}
  {\bibinfo  {journal} {Proceedings of the IEEE}\ }\textbf {\bibinfo {volume}
  {51}},\ \bibinfo {pages} {89} (\bibinfo {year} {1963})},\ \bibinfo {note}
  {04773}\BibitemShut {NoStop}%
\bibitem [{\citenamefont {Forn-D{\'\i}az}\ \emph {et~al.}(2010)\citenamefont
  {Forn-D{\'\i}az}, \citenamefont {Lisenfeld}, \citenamefont {Marcos},
  \citenamefont {Garc{\'\i}a-Ripoll}, \citenamefont {Solano}, \citenamefont
  {Harmans},\ and\ \citenamefont {Mooij}}]{forn2010observation}%
  \BibitemOpen
  \bibfield  {author} {\bibinfo {author} {\bibfnamefont {P.}~\bibnamefont
  {Forn-D{\'\i}az}}, \bibinfo {author} {\bibfnamefont {J.}~\bibnamefont
  {Lisenfeld}}, \bibinfo {author} {\bibfnamefont {D.}~\bibnamefont {Marcos}},
  \bibinfo {author} {\bibfnamefont {J.~J.}\ \bibnamefont {Garc{\'\i}a-Ripoll}},
  \bibinfo {author} {\bibfnamefont {E.}~\bibnamefont {Solano}}, \bibinfo
  {author} {\bibfnamefont {C.~J. P.~M.}\ \bibnamefont {Harmans}}, \ and\
  \bibinfo {author} {\bibfnamefont {J.~E.}\ \bibnamefont {Mooij}},\ }\href@noop
  {} {\bibfield  {journal} {\bibinfo  {journal} {Phys. Rev. Lett.}\ }\textbf
  {\bibinfo {volume} {105}},\ \bibinfo {pages} {237001} (\bibinfo {year}
  {2010})}\BibitemShut {NoStop}%
\bibitem [{\citenamefont {Singh}\ \emph {et~al.}(2014)\citenamefont {Singh},
  \citenamefont {Schneider}, \citenamefont {Bosman}, \citenamefont {Merkx},\
  and\ \citenamefont {Steele}}]{singh2014molybdenum}%
  \BibitemOpen
  \bibfield  {author} {\bibinfo {author} {\bibfnamefont {V.}~\bibnamefont
  {Singh}}, \bibinfo {author} {\bibfnamefont {B.~H.}\ \bibnamefont
  {Schneider}}, \bibinfo {author} {\bibfnamefont {S.~J.}\ \bibnamefont
  {Bosman}}, \bibinfo {author} {\bibfnamefont {E.~P.~J.}\ \bibnamefont
  {Merkx}}, \ and\ \bibinfo {author} {\bibfnamefont {G.~A.}\ \bibnamefont
  {Steele}},\ }\href@noop {} {\bibfield  {journal} {\bibinfo  {journal} {Appl.
  Phys. Lett.}\ }\textbf {\bibinfo {volume} {105}},\ \bibinfo {pages} {222601}
  (\bibinfo {year} {2014})}\BibitemShut {NoStop}%
\bibitem [{\citenamefont {Bishop}\ \emph {et~al.}(2010)\citenamefont {Bishop},
  \citenamefont {Ginossar},\ and\ \citenamefont {Girvin}}]{bishop2010response}%
  \BibitemOpen
  \bibfield  {author} {\bibinfo {author} {\bibfnamefont {L.~S.}\ \bibnamefont
  {Bishop}}, \bibinfo {author} {\bibfnamefont {E.}~\bibnamefont {Ginossar}}, \
  and\ \bibinfo {author} {\bibfnamefont {S.~M.}\ \bibnamefont {Girvin}},\
  }\href@noop {} {\bibfield  {journal} {\bibinfo  {journal} {Phys. Rev. Lett.}\
  }\textbf {\bibinfo {volume} {105}},\ \bibinfo {pages} {100505} (\bibinfo
  {year} {2010})}\BibitemShut {NoStop}%
\bibitem [{\citenamefont {Fink}\ \emph {et~al.}(2010)\citenamefont {Fink},
  \citenamefont {Steffen}, \citenamefont {Studer}, \citenamefont {Bishop},
  \citenamefont {Baur}, \citenamefont {Bianchetti}, \citenamefont {Bozyigit},
  \citenamefont {Lang}, \citenamefont {Filipp}, \citenamefont {Leek},\ and\
  \citenamefont {Wallraff}}]{fink2010quantum}%
  \BibitemOpen
  \bibfield  {author} {\bibinfo {author} {\bibfnamefont {J.~M.}\ \bibnamefont
  {Fink}}, \bibinfo {author} {\bibfnamefont {L.}~\bibnamefont {Steffen}},
  \bibinfo {author} {\bibfnamefont {P.}~\bibnamefont {Studer}}, \bibinfo
  {author} {\bibfnamefont {L.~S.}\ \bibnamefont {Bishop}}, \bibinfo {author}
  {\bibfnamefont {M.}~\bibnamefont {Baur}}, \bibinfo {author} {\bibfnamefont
  {R.}~\bibnamefont {Bianchetti}}, \bibinfo {author} {\bibfnamefont
  {D.}~\bibnamefont {Bozyigit}}, \bibinfo {author} {\bibfnamefont
  {C.}~\bibnamefont {Lang}}, \bibinfo {author} {\bibfnamefont {S.}~\bibnamefont
  {Filipp}}, \bibinfo {author} {\bibfnamefont {P.~J.}\ \bibnamefont {Leek}}, \
  and\ \bibinfo {author} {\bibfnamefont {A.}~\bibnamefont {Wallraff}},\
  }\href@noop {} {\bibfield  {journal} {\bibinfo  {journal} {Phys. Rev. Lett.}\
  }\textbf {\bibinfo {volume} {105}},\ \bibinfo {pages} {163601} (\bibinfo
  {year} {2010})}\BibitemShut {NoStop}%
\bibitem [{\citenamefont {Wallraff}\ \emph {et~al.}(2004)\citenamefont
  {Wallraff}, \citenamefont {Schuster}, \citenamefont {Blais}, \citenamefont
  {Frunzio}, \citenamefont {Huang}, \citenamefont {Majer}, \citenamefont
  {Kumar}, \citenamefont {Girvin},\ and\ \citenamefont
  {Schoelkopf}}]{wallraff2004strong}%
  \BibitemOpen
  \bibfield  {author} {\bibinfo {author} {\bibfnamefont {A.}~\bibnamefont
  {Wallraff}}, \bibinfo {author} {\bibfnamefont {D.~I.}\ \bibnamefont
  {Schuster}}, \bibinfo {author} {\bibfnamefont {A.}~\bibnamefont {Blais}},
  \bibinfo {author} {\bibfnamefont {L.}~\bibnamefont {Frunzio}}, \bibinfo
  {author} {\bibfnamefont {R.-S.}\ \bibnamefont {Huang}}, \bibinfo {author}
  {\bibfnamefont {J.}~\bibnamefont {Majer}}, \bibinfo {author} {\bibfnamefont
  {S.}~\bibnamefont {Kumar}}, \bibinfo {author} {\bibfnamefont {S.~M.}\
  \bibnamefont {Girvin}}, \ and\ \bibinfo {author} {\bibfnamefont {R.~J.}\
  \bibnamefont {Schoelkopf}},\ }\href@noop {} {\bibfield  {journal} {\bibinfo
  {journal} {Nature}\ }\textbf {\bibinfo {volume} {431}},\ \bibinfo {pages}
  {162} (\bibinfo {year} {2004})}\BibitemShut {NoStop}%
\bibitem [{\citenamefont {Samkharadze}\ \emph {et~al.}(2016)\citenamefont
  {Samkharadze}, \citenamefont {Bruno}, \citenamefont {Scarlino}, \citenamefont
  {Zheng}, \citenamefont {DiVincenzo}, \citenamefont {DiCarlo},\ and\
  \citenamefont {Vandersypen}}]{samkharadze2016high}%
  \BibitemOpen
  \bibfield  {author} {\bibinfo {author} {\bibfnamefont {N.}~\bibnamefont
  {Samkharadze}}, \bibinfo {author} {\bibfnamefont {A.}~\bibnamefont {Bruno}},
  \bibinfo {author} {\bibfnamefont {P.}~\bibnamefont {Scarlino}}, \bibinfo
  {author} {\bibfnamefont {G.}~\bibnamefont {Zheng}}, \bibinfo {author}
  {\bibfnamefont {D.~P.}\ \bibnamefont {DiVincenzo}}, \bibinfo {author}
  {\bibfnamefont {L.}~\bibnamefont {DiCarlo}}, \ and\ \bibinfo {author}
  {\bibfnamefont {L.~M.~K.}\ \bibnamefont {Vandersypen}},\ }\href@noop {}
  {\bibfield  {journal} {\bibinfo  {journal} {Phys. Rev. Applied}\ }\textbf
  {\bibinfo {volume} {5}},\ \bibinfo {pages} {044004} (\bibinfo {year}
  {2016})}\BibitemShut {NoStop}%
\bibitem [{\citenamefont {Andersen}\ and\ \citenamefont
  {Blais}(2016)}]{andersen2016ultrastrong}%
  \BibitemOpen
  \bibfield  {author} {\bibinfo {author} {\bibfnamefont {C.~K.}\ \bibnamefont
  {Andersen}}\ and\ \bibinfo {author} {\bibfnamefont {A.}~\bibnamefont
  {Blais}},\ }\href@noop {} {\bibfield  {journal} {\bibinfo  {journal} {arXiv
  preprint arXiv:1607.03770}\ } (\bibinfo {year} {2016})}\BibitemShut {NoStop}%
\end{thebibliography}%


%merlin.mbs apsrev4-1.bst 2010-07-25 4.21a (PWD, AO, DPC) hacked
%Control: key (0)
%Control: author (8) initials jnrlst
%Control: editor formatted (1) identically to author
%Control: production of article title (-1) disabled
%Control: page (0) single
%Control: year (1) truncated
%Control: production of eprint (0) enabled
\providecommand{\noopsort}[1]{}\providecommand{\singleletter}[1]{#1}%
\begin{thebibliography}{6}%
\makeatletter
\providecommand \@ifxundefined [1]{%
 \@ifx{#1\undefined}
}%
\providecommand \@ifnum [1]{%
 \ifnum #1\expandafter \@firstoftwo
 \else \expandafter \@secondoftwo
 \fi
}%
\providecommand \@ifx [1]{%
 \ifx #1\expandafter \@firstoftwo
 \else \expandafter \@secondoftwo
 \fi
}%
\providecommand \natexlab [1]{#1}%
\providecommand \enquote  [1]{``#1''}%
\providecommand \bibnamefont  [1]{#1}%
\providecommand \bibfnamefont [1]{#1}%
\providecommand \citenamefont [1]{#1}%
\providecommand \href@noop [0]{\@secondoftwo}%
\providecommand \href [0]{\begingroup \@sanitize@url \@href}%
\providecommand \@href[1]{\@@startlink{#1}\@@href}%
\providecommand \@@href[1]{\endgroup#1\@@endlink}%
\providecommand \@sanitize@url [0]{\catcode `\\12\catcode `\$12\catcode
  `\&12\catcode `\#12\catcode `\^12\catcode `\_12\catcode `\%12\relax}%
\providecommand \@@startlink[1]{}%
\providecommand \@@endlink[0]{}%
\providecommand \url  [0]{\begingroup\@sanitize@url \@url }%
\providecommand \@url [1]{\endgroup\@href {#1}{\urlprefix }}%
\providecommand \urlprefix  [0]{URL }%
\providecommand \Eprint [0]{\href }%
\providecommand \doibase [0]{http://dx.doi.org/}%
\providecommand \selectlanguage [0]{\@gobble}%
\providecommand \bibinfo  [0]{\@secondoftwo}%
\providecommand \bibfield  [0]{\@secondoftwo}%
\providecommand \translation [1]{[#1]}%
\providecommand \BibitemOpen [0]{}%
\providecommand \bibitemStop [0]{}%
\providecommand \bibitemNoStop [0]{.\EOS\space}%
\providecommand \EOS [0]{\spacefactor3000\relax}%
\providecommand \BibitemShut  [1]{\csname bibitem#1\endcsname}%
\let\auto@bib@innerbib\@empty
%</preamble>
\bibitem [{\citenamefont {Bosman}\ \emph {et~al.}(2015)\citenamefont {Bosman},
  \citenamefont {Singh}, \citenamefont {Bruno},\ and\ \citenamefont
  {Steele}}]{bosman2015broadband}%
  \BibitemOpen
  \bibfield  {author} {\bibinfo {author} {\bibfnamefont {S.~J.}\ \bibnamefont
  {Bosman}}, \bibinfo {author} {\bibfnamefont {V.}~\bibnamefont {Singh}},
  \bibinfo {author} {\bibfnamefont {A.}~\bibnamefont {Bruno}}, \ and\ \bibinfo
  {author} {\bibfnamefont {G.~A.}\ \bibnamefont {Steele}},\ }\href@noop {}
  {\bibfield  {journal} {\bibinfo  {journal} {Appl. Phys. Lett.}\ }\textbf
  {\bibinfo {volume} {107}},\ \bibinfo {pages} {192602} (\bibinfo {year}
  {2015})}\BibitemShut {NoStop}%
\bibitem [{\citenamefont {Gely}\ \emph {et~al.}(2017)\citenamefont {Gely},
  \citenamefont {Parra-Rodriguez}, \citenamefont {Bothner}, \citenamefont
  {Blanter}, \citenamefont {Bosman}, \citenamefont {Solano},\ and\
  \citenamefont {Steele}}]{arxiv_gely_divergence-free_2017}%
  \BibitemOpen
  \bibfield  {author} {\bibinfo {author} {\bibfnamefont {M.~F.}\ \bibnamefont
  {Gely}}, \bibinfo {author} {\bibfnamefont {A.}~\bibnamefont
  {Parra-Rodriguez}}, \bibinfo {author} {\bibfnamefont {D.}~\bibnamefont
  {Bothner}}, \bibinfo {author} {\bibfnamefont {Y.~M.}\ \bibnamefont
  {Blanter}}, \bibinfo {author} {\bibfnamefont {S.~J.}\ \bibnamefont {Bosman}},
  \bibinfo {author} {\bibfnamefont {E.}~\bibnamefont {Solano}}, \ and\ \bibinfo
  {author} {\bibfnamefont {G.~A.}\ \bibnamefont {Steele}},\ }\href@noop {}
  {\bibfield  {journal} {\bibinfo  {journal} {arXiv:1701.05095}\ } (\bibinfo
  {year} {2017})}\BibitemShut {NoStop}%
\bibitem [{\citenamefont {Koch}\ \emph {et~al.}(2007)\citenamefont {Koch},
  \citenamefont {Yu}, \citenamefont {Gambetta}, \citenamefont {Houck},
  \citenamefont {Schuster}, \citenamefont {Majer}, \citenamefont {Blais},
  \citenamefont {Devoret}, \citenamefont {Girvin},\ and\ \citenamefont
  {Schoelkopf}}]{koch2007charge}%
  \BibitemOpen
  \bibfield  {author} {\bibinfo {author} {\bibfnamefont {J.}~\bibnamefont
  {Koch}}, \bibinfo {author} {\bibfnamefont {T.~M.}\ \bibnamefont {Yu}},
  \bibinfo {author} {\bibfnamefont {J.}~\bibnamefont {Gambetta}}, \bibinfo
  {author} {\bibfnamefont {A.~A.}\ \bibnamefont {Houck}}, \bibinfo {author}
  {\bibfnamefont {D.~I.}\ \bibnamefont {Schuster}}, \bibinfo {author}
  {\bibfnamefont {J.}~\bibnamefont {Majer}}, \bibinfo {author} {\bibfnamefont
  {A.}~\bibnamefont {Blais}}, \bibinfo {author} {\bibfnamefont {M.~H.}\
  \bibnamefont {Devoret}}, \bibinfo {author} {\bibfnamefont {S.~M.}\
  \bibnamefont {Girvin}}, \ and\ \bibinfo {author} {\bibfnamefont {R.~J.}\
  \bibnamefont {Schoelkopf}},\ }\href@noop {} {\bibfield  {journal} {\bibinfo
  {journal} {Phys. Rev. A}\ }\textbf {\bibinfo {volume} {76}},\ \bibinfo
  {pages} {042319} (\bibinfo {year} {2007})}\BibitemShut {NoStop}%
\bibitem [{\citenamefont {Bishop}\ \emph {et~al.}(2010)\citenamefont {Bishop},
  \citenamefont {Ginossar},\ and\ \citenamefont {Girvin}}]{bishop2010response}%
  \BibitemOpen
  \bibfield  {author} {\bibinfo {author} {\bibfnamefont {L.~S.}\ \bibnamefont
  {Bishop}}, \bibinfo {author} {\bibfnamefont {E.}~\bibnamefont {Ginossar}}, \
  and\ \bibinfo {author} {\bibfnamefont {S.~M.}\ \bibnamefont {Girvin}},\
  }\href@noop {} {\bibfield  {journal} {\bibinfo  {journal} {Phys. Rev. Lett.}\
  }\textbf {\bibinfo {volume} {105}},\ \bibinfo {pages} {100505} (\bibinfo
  {year} {2010})}\BibitemShut {NoStop}%
\bibitem [{\citenamefont {Schreier}\ \emph {et~al.}(2008)\citenamefont
  {Schreier}, \citenamefont {Houck}, \citenamefont {Koch}, \citenamefont
  {Schuster}, \citenamefont {Johnson}, \citenamefont {Chow}, \citenamefont
  {Gambetta}, \citenamefont {Majer}, \citenamefont {Frunzio}, \citenamefont
  {Devoret}, \citenamefont {Girvin},\ and\ \citenamefont
  {Schoelkopf}}]{schreier2008suppressing}%
  \BibitemOpen
  \bibfield  {author} {\bibinfo {author} {\bibfnamefont {J.}~\bibnamefont
  {Schreier}}, \bibinfo {author} {\bibfnamefont {A.~A.}\ \bibnamefont {Houck}},
  \bibinfo {author} {\bibfnamefont {J.}~\bibnamefont {Koch}}, \bibinfo {author}
  {\bibfnamefont {D.~I.}\ \bibnamefont {Schuster}}, \bibinfo {author}
  {\bibfnamefont {B.}~\bibnamefont {Johnson}}, \bibinfo {author} {\bibfnamefont
  {J.}~\bibnamefont {Chow}}, \bibinfo {author} {\bibfnamefont {J.~M.}\
  \bibnamefont {Gambetta}}, \bibinfo {author} {\bibfnamefont {J.}~\bibnamefont
  {Majer}}, \bibinfo {author} {\bibfnamefont {L.}~\bibnamefont {Frunzio}},
  \bibinfo {author} {\bibfnamefont {M.~H.}\ \bibnamefont {Devoret}}, \bibinfo
  {author} {\bibfnamefont {S.~M.}\ \bibnamefont {Girvin}}, \ and\ \bibinfo
  {author} {\bibfnamefont {R.~J.}\ \bibnamefont {Schoelkopf}},\ }\href@noop {}
  {\bibfield  {journal} {\bibinfo  {journal} {Phys. Rev. B}\ }\textbf {\bibinfo
  {volume} {77}},\ \bibinfo {pages} {180502} (\bibinfo {year}
  {2008})}\BibitemShut {NoStop}%
\bibitem [{\citenamefont {Schuster}(2007)}]{schuster_circuit_2007}%
  \BibitemOpen
  \bibfield  {author} {\bibinfo {author} {\bibfnamefont {D.~I.}\ \bibnamefont
  {Schuster}},\ }\href@noop {} {\emph {\bibinfo {title} {Circuit quantum
  electrodynamics}}}\ (\bibinfo  {publisher} {Yale University},\ \bibinfo
  {year} {2007})\BibitemShut {NoStop}%
\end{thebibliography}%

\end{document}

\title{Multi-mode ultra-strong coupling in circuit quantum electrodynamics:\\
Supplementary material}% Force line breaks with \\
%\thanks{A footnote to the article title}%

\author{Sal J. Bosman}
% \altaffiliation[Also at ]{Physics Department, XYZ University.}%Lines break automatically or can be forced with \\
\affiliation{%
Kavli Institute of NanoScience, Delft University of Technology,\\
PO Box 5046, 2600 GA, Delft, The Netherlands.
% This line break forced with \textbackslash\textbackslash
}%
%  \email{s.j.bosman@tudelft.nl}

\author{Mario F. Gely}
% \altaffiliation[Also at ]{Physics Department, XYZ University.}%Lines break automatically or can be forced with \\
\affiliation{%
Kavli Institute of NanoScience, Delft University of Technology,\\
PO Box 5046, 2600 GA, Delft, The Netherlands.
% This line break forced with \textbackslash\textbackslash
}%
%  \email{s.j.bosman@tudelft.nl}
\author{Vibhor Singh}%
 \affiliation{%
Department of Physics, Indian Institute of Science, Bangalore 560012, India
}%

\author{Alessandro Bruno}
\affiliation{ 
Qutech Advanced Research Center, Delft University of Technology,\\
Lorentzweg 1, 2628 CJ Delft, The Netherlands.}

\author{Daniel Bothner}%
 \affiliation{%
Kavli Institute of NanoScience, Delft University of Technology,\\
PO Box 5046, 2600 GA, Delft, The Netherlands.
% This line break forced with \textbackslash\textbackslash
}%

\author{Gary A. Steele}

\affiliation{%
Kavli Institute of NanoScience, Delft University of Technology,\\
PO Box 5046, 2600 GA, Delft, The Netherlands.
% This line break forced with \textbackslash\textbackslash
}%

\maketitle

\renewcommand{\theequation}{S\arabic{equation}}
\renewcommand{\thefigure}{S\arabic{figure}}
\renewcommand{\thetable}{S\arabic{table}}

\renewcommand{\thesection}{S\arabic{section}}
\renewcommand{\thesubsection}{S\arabic{subsection}}
\renewcommand{\thesubsubsection}{S\arabic{subsubsection}}
\renewcommand{\bibnumfmt}[1]{[S#1]}
\renewcommand{\citenumfont}[1]{S#1}

\onecolumngrid
\clearpage
\FloatBarrier\section{Measurement setup}

\begin{figure}[b!]
\includegraphics[width=9.5cm]{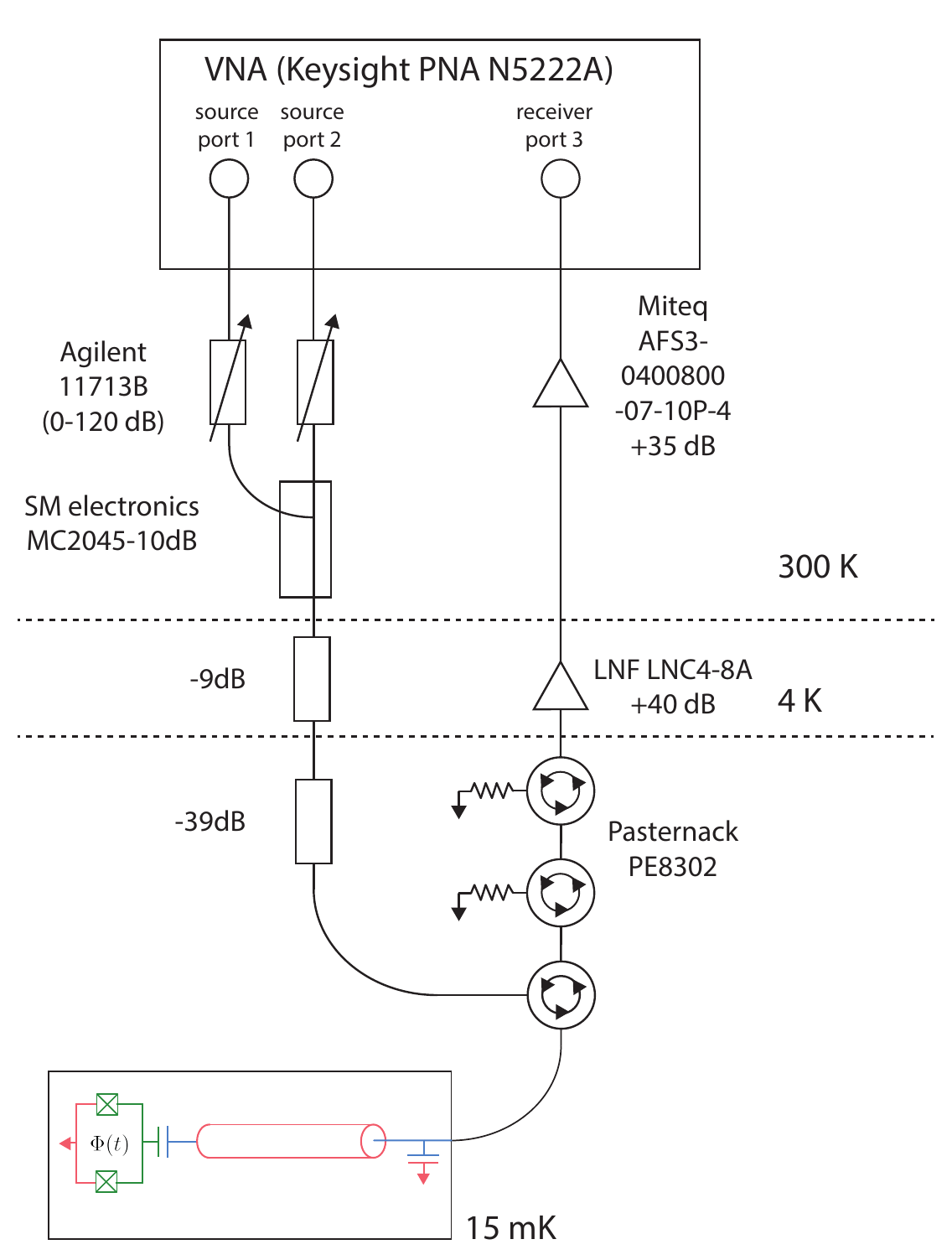}
\caption{\label{fig:wide} Schematic of the measurement setup.  
}
\label{fig:setup}
\end{figure}

Supplementary Fig.~\ref{fig:setup} shows schematically the measurement setup used for the device characterization. The vector network analyser (VNA) outputs one or two continious wave (CW) signals that are sent through a variable attenuator (0-120 dB) and combined with a directional coupler. From there the signal is sent into the dilution fridge, where it is attenuated (48 dB) before reaching the sample through a circulator. The reflected signal from the device is sent back to the VNA using two isolators and amplifiers. 

\FloatBarrier\section{Supplementary Spectra}

\begin{figure}[b!]
\includegraphics[width=16cm]{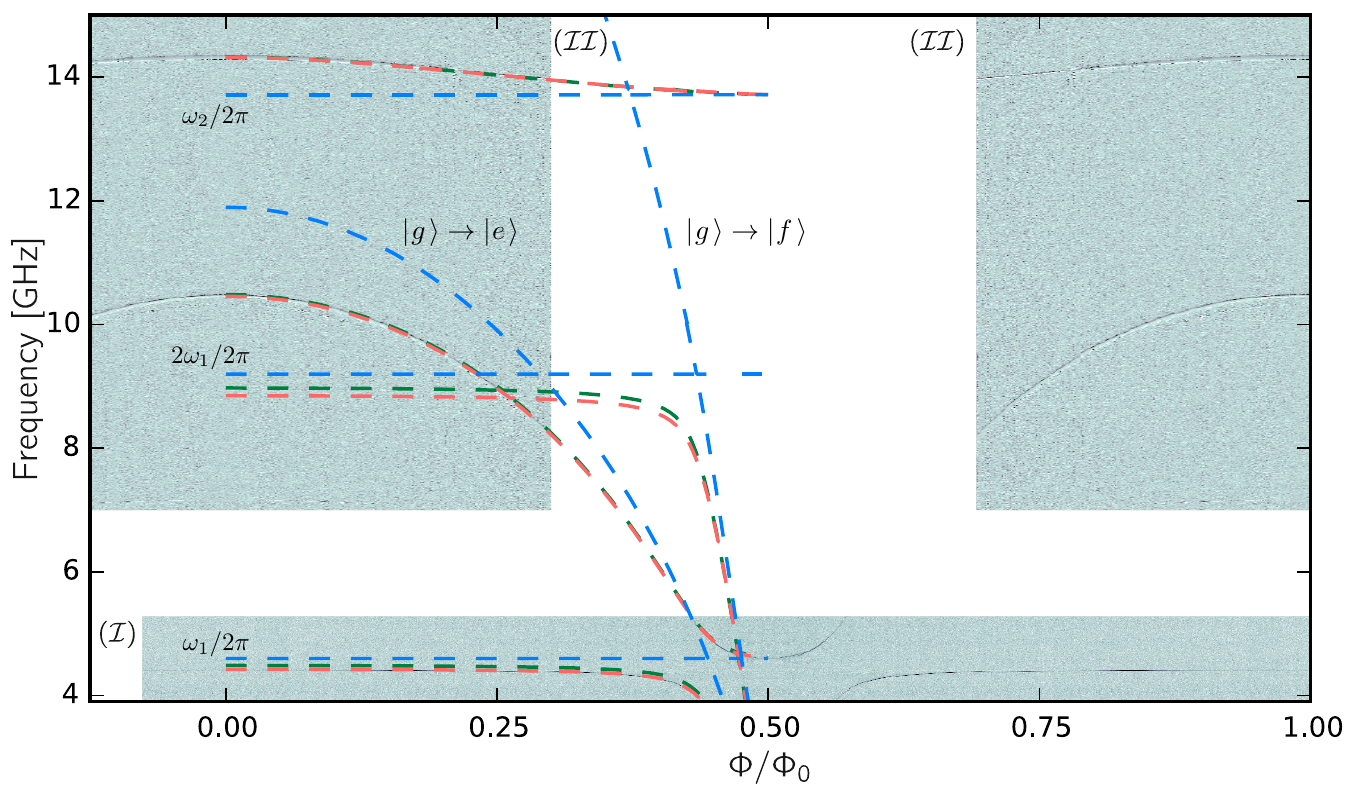}
\caption{\label{fig:wide} Spectrum of device A, as represented as Fig.\@ 3 in the main text for device B. 
}
\label{fig:spec_A}
\end{figure}

Supplementary Fig.~\ref{fig:spec_A} shows the spectrum of device A (155 nm vacuum gap capacitor). The maximum frequency of the qubit is $\sim 12$ GHz, and the first harmonic of the cavity is at $\sim 14$ GHz, leading to only a single vacuum Rabi splitting in this device. The second energy level of the fundamental mode of the resonator is indicated in the blue dashed lines (labeled $2\omega_1/2\pi$), as the corresponding state hybridizes with the $\ket{g} \rightarrow \ket{f}$ transition we observed in Fig.\@ 2(c) in the main text.  \\

\begin{figure}[h!]
\includegraphics[width=8cm]{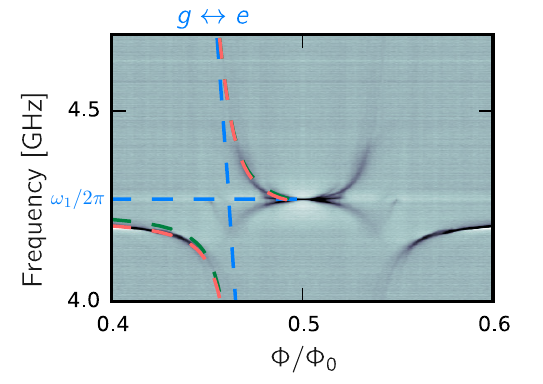}
\caption{\label{fig:wide} Single-tone spectrum of device B, as represented as Fig.\@ 2 in the main text for device A.
}
\label{fig:spec_C}
\end{figure}

% \vspace{70pt}
Supplementary Fig.~\ref{fig:spec_C} shows the response of device B (350 nm vacuum gap capacitor) in single-tone microwave reflectometry. From our model, we can attribute the faint lines bending downwards (opposite to the upper branch of the VRS mirrored in the line $\omega_1$) to the dressed frequency of the resonator with the Cooper pair box in the excited state $|e\rangle$ (in this flux region the Josephson energy is too small to call the qubit a Transmon). The response of the circuit also goes to the bare resonator frequency at half a flux quantum indicating that the junctions are also very symmetric in this device. 

\FloatBarrier\section{Model}

\begin{figure}[h!]
  \centering
  \includegraphics[width=80mm]{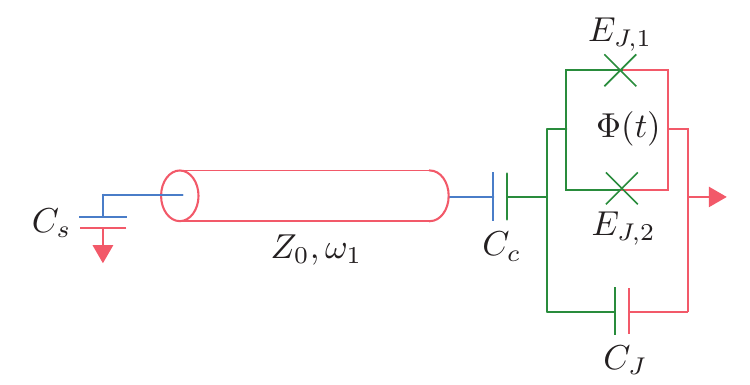}
  \caption{Complete circuit model of our device. The Transmon is a charge island connected to ground through a flux-biased SQUID and a capacitance $C_J$ and connected to a $\lambda/4$ resonator through a capacitance $C_c$.}
  \label{fig:circuit}
 
\end{figure}

In this section we discuss the modeling of the device shown schematically in Fig.~\ref{fig:circuit}. In the GHz regime we are working in ($\omega/2\pi>3.5$ GHz), the impedance of the $C_s=30$ pF shunt capacitor is small ($|1/i\omega C_{s}|\simeq2\ \Omega\ll Z_0=50\ \Omega$) such that it can effectively be considered as a short to ground~\cite{bosman2015broadband}. We can then use the model derived in the associated theoretical paper~\cite{arxiv_gely_divergence-free_2017}, adding however two elements. First, the flux dependent Josephson term (see Ref. \cite{koch2007charge})
  \begin{equation}
  - E_J \cos (\hat{\delta})\rightarrow- E_J \cos  (\hat{\delta})\cos \bigg(\pi\frac{\Phi}{\Phi_0}\bigg) - dE_J \sin (\hat{\delta})\sin \bigg(\pi\frac{\Phi}{\Phi_0}\bigg)\ ,
  \label{eq:Ej_flux_dep}
  \end{equation}
 where $\hat{\delta}$ is the superconducting phase difference across the SQUID. We defined the total Josephson energy by $E_J = E_{J,1} + E_{J,2}$ and the asymmetry by $d =( E_{J,1} - E_{J,2})/(E_{J,1} + E_{J,2})$. Secondly, we offset the quantum number of Cooper pairs on the Transmon island by a constant value $n_\text{DC}$ to model an environmental offset charge present in any realistic system in the Cooper pair box regime~\cite{koch2007charge}.

The Hamiltonian diagonalization is performed in two steps, first a diagonalization of the Cooper pair box Hamiltonian in the charge basis, secondly of an extended Rabi Hamiltonian. For more details, see the supplementary material of Ref.\cite{arxiv_gely_divergence-free_2017}. In order to perform numerics, four simulation parameters should be fixed:
\begin{enumerate}
  \item A maximum number of Cooper pairs on the Transmons charge island $N_\text{max}$
  \item The number of Transmon levels $N_q$
  \item The number of resonator modes $N_\text{modes}$...
  \item ...each with a certain number of photons $\{N_{m}\}_{m = 0,..,N_\text{modes}-1}$
\end{enumerate}

 The size of the Hilbert space used for the diagonalization of the Cooper pair box Hamiltonian scales with $2N_\text{max}+1$ and for the diagonalization of the extended Rabi Hamiltonian it scales with $2^{N_q}\prod N_{m}$. We should therefore have a small enough Hilbert space such that the diagonalizations are feasible with the computer resources at our disposal, whilst ensuring that the neglected degrees of freedom do not significantly change the spectrum if we would have included them. We consider that a degree of freedom which changes the computed spectrum by less than a tenth of the measured line-width can be neglected. This condition leads to the following simulation parameters $N_\text{max} = 20,N_\text{modes}=4,N_q=5$. The number of photon levels to include depends on the strength of the coupling and we have used 6,4,3 and 3 (5,3,2 and 2) photon levels in the fundamental, first, second and third modes of the resonator for device A (B)

\FloatBarrier\section{Fitting}
\subsection{Fitting routine}

We first experimentally perform a broad flux-dependent single tone measurement of the dressed cavity in order to extract the current periodicity and the point of maximum qubit frequency. This allows us to convert current (that we apply to a coil to bias the SQUID) to the flux through the SQUID (in units of flux quantum). Nine free parameters are to be determined for both devices: $\omega_1, C_c, C_J, n_\text{DC}, E_J, d$. Due to the effect of the shunt capacitor (neglected in the model) we also correct the frequency of the higher modes leading to three more parameters: $\alpha_n$, such that $\omega_m = \alpha_m(2m-1)\omega_1$, $m\ge1$. We expect $\alpha_m\simeq1$ and shall maintain this hypothesis until the last stage of the fit. Two other parameters can be fixed easily:
\begin{enumerate}
  \item Through a high power single-tone measurement, we can drive the (dressed) fundamental mode of the resonator to its harmonic regime \cite{bishop2010response} and determine the bare resonator frequency $\omega_1$.
  \item For both devices, at a flux of $\Phi_0/2$, the dressed cavity frequency resumes its bare frequency which indicates that the coupling is negligible at that flux point. In simulation, this can only be achieved for very low asymmetry $d\simeq0.01$.
\end{enumerate}

We then perform a least-square minimization routine, fitting the extracted data -- the dressed first and second cavity mode, and dressed first Transmon transition frequency, as a function of flux -- to a diagonalization of the Hamiltonian. The free parameters in this routine are $C_c$, $C_J$, $n_\text{DC}$, $E_J$. Due to the considerable time necessary to perform the diagonalization (approximatively ten seconds per flux point on a commercial laptop computer), we parallelize the diagonalization over the different flux points using a high performance computing cluster. Finally we adjust the frequencies of the higher modes through the parameters $\alpha_m$.

\subsection{Differentiating $E_c$ and $E_J$}

\begin{figure}[h!]
  \centering
  \includegraphics[width=90mm]{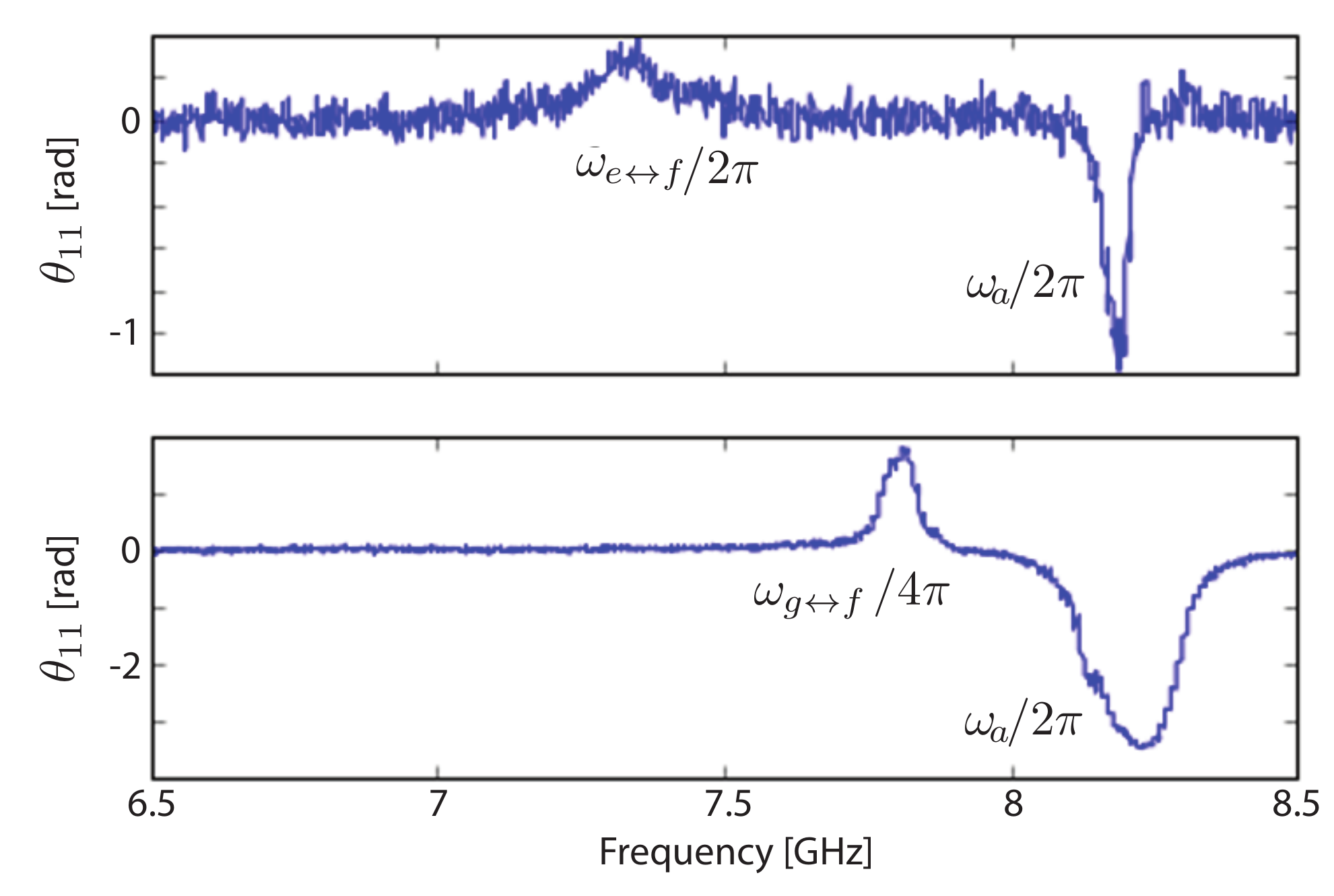}
  \caption{In the top panel we measure the excited to second excited transition frequency of the Transmon by performing two-tone spectroscopy whilst exiting the qubit with a third drive tone. In the lower panel we perform a similar experiment however substituting the third tone with high probe tone powers. On the $y$-axis we plot $\theta_{11}(\omega)$ the phase of the cavity response at $\omega_1$, denoted by $\tilde{S}_{11}(\omega)$ in the main text. This allows us to measure the two-photon process which excites the second excited Transmon state from the ground state. The transition frequencies are shifted and broadened for strong probe powers (observed previously, for example in \cite{schuster_circuit_2007} Fig.~8.7), but the measured anharmonicity is constant. The two measurements where performed at the same flux point (0.3752 flux quantum). They both provide a measurement of the anharmonicity and the low power three-tone measurement also provides an exact measurement of the ground to excited and second excited state at that particular flux point.}
  \label{fig:Ec}
 
\end{figure}

Since the frequency of the first Transmon transition is approximated by $\hbar\omega_{a} \simeq \sqrt{8E_cE_J}-E_c$, simply fitting the dressed first Transmon transition leads to an imprecision in the estimation of $E_c$ and $E_J$. Namely different combinations of $E_c$ and $E_J$ lead to good fits to the data. In order to lift this indeterminacy, we measure the second level of the Transmon (and thus the anharmonicity of the device) and insure that the fitted parameters accurately predict its dressed frequency. We used this approach for device B (see fig. \ref{fig:Ec}) reproducing previously observed results \cite{schreier2008suppressing}.\\

We do not have a measurement of the second Transmon transition of device A (now un-measurable). The two SQUIDs where however fabricated on the same wafer, simultaneously, and with the same design parameters. We will therefore consider that the Josephson energies of the two devices should be equal. Whilst this fact weakens the predictive power of our fit of device A, it does not impact our estimation of the coupling.

\subsection{Fitted Device Parameters}

In supplementary tables S1, S2 and S3 we tabulate all the device characteristics, either fixed or extracted from the fit of our model as described above.
\begin{table}[h]
    \begin{tabular} {lllll}%{|c|c||>{\centering}p{.9cm}|>{\centering}p{.9cm}|>{\centering}p{.9cm}|c|}
        \hline
        Quantity & Symbol  & Device A & Device B  & unit \\ \hline
        \textbf{Resonator parameters} & & & \\ 
        bare fundamental frequency & $\omega_1/2\pi$  & 4.603 & 4.268 & GHz \\

        %linewidth fundamental mode at 0 flux & $\kappa_1(0)$ [MHz] & & \\
        %bare second mode frequency & $\omega_2$ [GHz] & 13.80 & 12.59 \\
        %linewidth second mode at 0 flux & $\kappa_2(0)$ [MHz] & & \\
        %bare third mode frequency & $\omega_3$ [GHz] & 13.80 & 12.59 \\
        %linewidth third mode at 0 flux & $\kappa_3(0)$ [MHz] & & \\
        %bare fourth mode frequency & $\omega_4$ [GHz] & 13.80 & 12.59 \\
        %linewidth fourth mode at 0 flux & $\kappa_4(0)$ [MHz] & & \\
        %bare fifth mode frequency & $\omega_5$ [GHz] & 13.80 & 12.59 \\
        %linewidth fifth mode at 0 flux & $\kappa_5(0)$ [MHz] & & \\
        %total length & $L$ & & & mm \\
        % total capacitance & $C_0$  & 541 & 583 & fF \\
        % total inductance & $L_0$  & 2.19 & 2.36 & nH \\
        resonator impedance & $Z_0$  & 50 & 50 & $\Omega$\\
        harmonics deviation & $[\alpha_m]_m$& [1.0,0.994,1.0,1.0] \vspace{15pt} & [1.0,0.983,0.983, 1.0]&\\ \hline
    \end{tabular} \\ \vskip .5cm

    \caption{Parameters of the $\lambda/4$ microwave resonators of the device.}
    \label{tab:truthTables}   
\end{table}
\begin{table}[h]
    \begin{tabular} {lllll}%{|c|c||>{\centering}p{.9cm}|>{\centering}p{.9cm}|>{\centering}p{.9cm}|c|}
        \hline
        Quantity & Symbol  & Device A & Device B & unit \\ \hline        
        \textbf{Transmon parameters} & & & \\
        charging energy & $E_c/h$  & 426 & 700 & MHz \\
        Josephson energy at 0 flux & $E_J(0)/h$  \hspace{5pt} & 36.3 & 37.4 & GHz \\
        Transmon parameter at 0 flux & $E_J(0)/E_c$ & 78.6 & 53.4 & \\
        %Transmon impedance at 0 flux & $Z_T$   & & & $\Omega$ \\
        maximum qubit frequency at 0 flux &$ \omega_{a}/2\pi $  & 10.67 & 13.74 & GHz \\ 
        Transmon capacitance to ground  &$C_J$  & 5.13 & 8.73 & fF\\
        vacuum-gap capacitor &$C_c$ & 40.3 & 18.9 & fF\\
        total capacitance & $C_\Sigma$  & 45.4 & 27.7& fF \\
        capacitance participation ratio & $\beta$ & 0.89 & 0.68 & \\
        distance of vacuum-gap capacitor & d & 155 & 330 & nm \\ \hline
        
    \end{tabular} \\ \vskip .5cm

    \caption{Transmon qubit parameters.}
    \label{tab:truthTables}   
\end{table}

\begin{table}[h]
    \begin{tabular} {lllll}%{|c|c||>{\centering}p{.9cm}|>{\centering}p{.9cm}|>{\centering}p{.9cm}|c|}
        \hline
        Quantity & Symbol  & Device A & Device B & unit \\ \hline        
        \textbf{circuit QED parameters} &  & & \\
        Rabi splitting with $\omega_1$ & $\Delta_1/2\pi$  &  1.19 & 0.63 & GHz \\
        Coupling ratio with $\omega_1$ & $\Delta_1/2\omega_1$  & 0.13 &  0.07 &  \\
        Rabi splitting with $\omega_2$ & $\Delta_2/2\pi$  & none & 1.82 & GHz \\
        Coupling ratio with $\omega_2$ & $\Delta_2/2\omega_2$  & none & 0.07 &  \\
        Coupling at 0 flux & $g(0)/2\pi$  & 897 & 568 & MHz \\
        Coupling ratio at 0 flux with $\omega_1$ & $g/\omega_1$ & 0.19 & 0.13 & \\
        %Maximal Bloch-Siegert shift $|1,-\rangle \otimes |0\rangle$ \hspace{10pt} & $\Sigma_{|1,-\rangle \otimes |0\rangle}$ & & &MHz \\
        %Maximal Bloch-Siegert shift $|0\rangle \otimes |1,-\rangle$ \hspace{10pt} & $\Sigma_{|0\rangle \otimes |1,-\rangle}$ & & &MHz \\
        %Maximal Bloch-Siegert shift $|0\rangle \otimes |1,+ \rangle$ \hspace{10pt} & $\Sigma_{|0\rangle \otimes |1,+\rangle}$ & & &MHz \\        
    \end{tabular} \\ \vskip .5cm

    \caption{Coupling parameters.}
    \label{tab:truthTables}   
\end{table}
\FloatBarrier
 \bibliography{library}